\documentclass[onecolumn]{aa}
\usepackage{graphicx}
\usepackage{txfonts}
\def\stdmdl{standard model }
\begin{document}

\title{Stochastic Processes, 
 Galactic Star Formation, and Chemical Evolution}
   \subtitle{Effects of Accretion, Stripping, and Collisions in
     Multiphase Multi-zone Models}

\titlerunning{Stochastic Processes, 
 Galactic Star Formation, and Chemical Evolution}

   \author{Giada Valle \inst{1}, Steven N. Shore \inst{1,2}, Daniele Galli\inst{3}}

\authorrunning{Giada Valle et al.}

   \institute{(1) Dipartimento di Fisica ``Enrico Fermi'', 
Universit\`a di Pisa, largo Pontecorvo 3, Pisa I-56127 Italy; \\
(2) INFN/Pisa, largo Pontecorvo 3, Pisa I-56127, Italy; \\
(3) INAF-Osservatorio Astrofisico di Arcetri, largo Enrico Fermi 5, 
Firenze I-50125 Italy}

   \offprints{Dr. G. Valle}

   \date{Received: 2/11/2004, Accepted 3/02/2005}

   \abstract{This paper reports simulations allowing for stochastic
accretion and mass loss  within closed and open systems modeled using 
a previously developed multi-population, multi-zone (halo, thick disk, 
thin disk) treatment.  The star formation rate is computed as a
function of time directly from the
model equations and all chemical evolution is followed 
without instantaneous recycling. 
Several types of simulations are presented here: (1) a closed system with 
bursty mass loss from the halo to the thick disk, and 
from the thick to the thin disk, in separate events to the thin disk; 
(2) open systems with 
random  environmental (extragalactic) accretion, {\it e.g}. by infall of
high velocity clouds directly to the thin disk; (3) 
schematic open system single and multiple collision events and intracluster
stripping.  For the open models, 
the mass of the Galaxy has been explicitly tracked with time.  We present the 
evolution of the star formation rate, metallicity histories, and concentrate 
on the light elements.  We find a wide range of possible outcomes, 
including an explanation for variations in the Galactic D/H ratio,  
and highlight the problems for uniquely reconstructing star forming 
histories from contemporary abundance measurements.

   }

   \maketitle

\section{Introduction}

The determination of galactic evolution over cosmic time requires  the
interpretation of the elemental abundance distribution as a  record of
changing star formation patterns.  Unraveling the   web of physical
processes has proven a challenge, complicated by the  many fundamental
uncertainties associated with specific scenarios  for the physical
processes.  A number of approaches, often complementary,  have been
tried.  These split broadly into  two categories.  One approach, {\it
Chemodynamics} include hydrodynamics (parameterized  chemical
evolution with either multiple fluids or SPH, etc.)  along with the
appropriate  nuclear yields but generally  use a simple formulation
for the star formation  rate as a function of gas density (see Hensler
\cite{hens03} and references therein).   The other uses analogs of
chemical networks and either assume a time  history for the star
formation rate in advance ({\it e.g.} Tinsley \cite{t80}, Hellsten \& 
Sommer-Larsen \cite{hsl92}, Pagel
\cite{pagel98}, Matteucci \cite{matt01}) or adopt a nonlinear
dynamical systems approach with  feedback  ({\it e.g.} Lynden-Bell
\cite{dlb75},  Shore \& Ferrini \cite{sf95}, Shore \& Franco
\cite{sfr00}).   Each has its successes in modeling some aspect(s) of
the system but most share  the feature that they are closed and
deterministic, conditions known to be respected more as 
exceptions than as a rule ({\it e.g.} 
van Gorkem \cite{vanG04}).  Only a few  
stochastic simulations have been attempted
date, especially  percolation models ({\it e.g.} Gerola \& Seiden \cite{ger78},
Matteucci \& Chiosi \cite{matt83}, Schulman \& Seiden \cite{schulman86}, Comins
\& Shore \cite{com90}, Perdang \& Lejeune \cite{per96})  and
fluctuating accretion  (Copi, Schramm, \& Turner \cite{copi95}, Copi
\cite{copi97}).

It is beyond the scope of this paper to review the  assumptions of
these individual approaches.  Instead, we will present  results of
ongoing simulations, using our previously developed  multi-population
framework ({\it cf.}  Ferrini et al. \cite{fmpp92}, Shore \& Ferrini
\cite{sf95}, Shore \& Galli \cite{sg03},  Valle et al. \cite{valle}),
that include stochastic  accretion  within closed and open systems
({\it cf.} Comins \& Shore \cite{com90}).  Our purpose is to show
that even these relatively schematic models, which have been recently
used to study the evolution of the light elements, display behavior
that  can inform more detailed modeling using any of the cited 
approaches of the dynamical and nucleosynthetic  processes 
associated with galactic evolution.

\section{Baseline Galactic chemical evolution model: the \stdmdl}

We begin with a description of our baseline, or {\it standard} model
that  has been developed for the Galaxy in a previous series of papers
so the changes we introduce here  will be understood in their proper
context. The model treats evolution within  a region formed by the
intersection with  the Galactic disk of a cylinder of radius $\sim 1$ kpc and
with its axis perpendicular to the disk.  The vertical height
corresponds to the scale length for the halo. The cylinder is centered
at the Solar galactocentric distance.  By the term {\it galaxy} we
mean only the actively star forming phase; a dark matter halo,
constituting the bulk of the mass, is assumed to be inertly present to
insure that there is no need to include dynamics explicitly in the
code.  Our technique  uses a multi-zone, multiphase approach.   Three
``zones'' (halo,  thick disk, and thin disk) of the galaxy  are
initialized with only the halo, for which only the diffuse gas phase
is  initially included and for which primordial abundances have been
assumed.  The other two form by accretion.   The time history of  star
formation is computed directly from the model equations rather than
being assumed {\it a priori} without instantaneous recycling and the
yields and deterministic rates are the  same as used in our previous
work (most recently, Valle et al. \cite{valle}).   We distinguish the
halo (HA), thick disk (TD) and thin disk (DI) and within each  include
three phases (plus an inert sink):  diffuse gas, clouds,
stars and remnants.  All  are expressed as the phase mass
fraction relative to the total mass of the system.  We assume that all
the matter is initially in the form of diffuse gas in halo, which
then begins to form the thick disk while forming clouds and then
stars.  Gas that falls onto the thick disk is already enriched from
nucleosynthesis by halo stars.   It forms clouds (through
gravitational instabilities, although  dynamical processes are not
included explicitly in this modeling scheme).   Star formation occurs
both  spontaneously, when stars are formed directly from the cloud
phase, and by stimulated processes through the interaction of clouds
and massive stars.  The same processes occur   in the thin disk.

The interactions among the different phases produce the time
dependence of the total mass fraction in each phase and the chemical
abundances in the interstellar medium (ISM) and in stars.  The
restitution of matter to the ISM is described by two time dependent
terms: the death rate $D(t)$ (the fraction of the total mass that
leaves the star phase at the time $t$) and the restitution rate $W(t)$
(the fraction of the mass injected in the interstellar gas from stars
dying at $t$). The computation of these terms is based on the
assumptions made for stellar evolution and nucleosynthesis (see
Ferrini et al. \cite{fmpp92}).  The star formation rate (SFR) is
determined consistently by the interaction among the different phases
of matter and no formal time dependence for the SFR is assumed {\em a
priori}.  The initial mass function (IMF) adopted in this model is
based on the  analysis of the fragmentation of molecular clouds in the
solar neighborhood (Ferrini et  al. 1991), and is constant in space
and time although we have tested the sensitivity of the final
abundances to this IMF by comparison of the mass averaged yields
using Kroupa (2002).  The principal effect of the mass distribution on
the final abundances is for elements whose formation has been delayed
by stellar evolution with respect to the stellar birth time, such as
Fe, since no instantaneous recycling approximation (IRA) is assumed in
our calculations.   The results, which will be reported in a future
paper, yielded no {\it qualitative} changes and thus, for
correspondence with our previously published work, we maintain the
original IMF here.

\begin{figure}
\centering
\includegraphics[width=14.5cm]{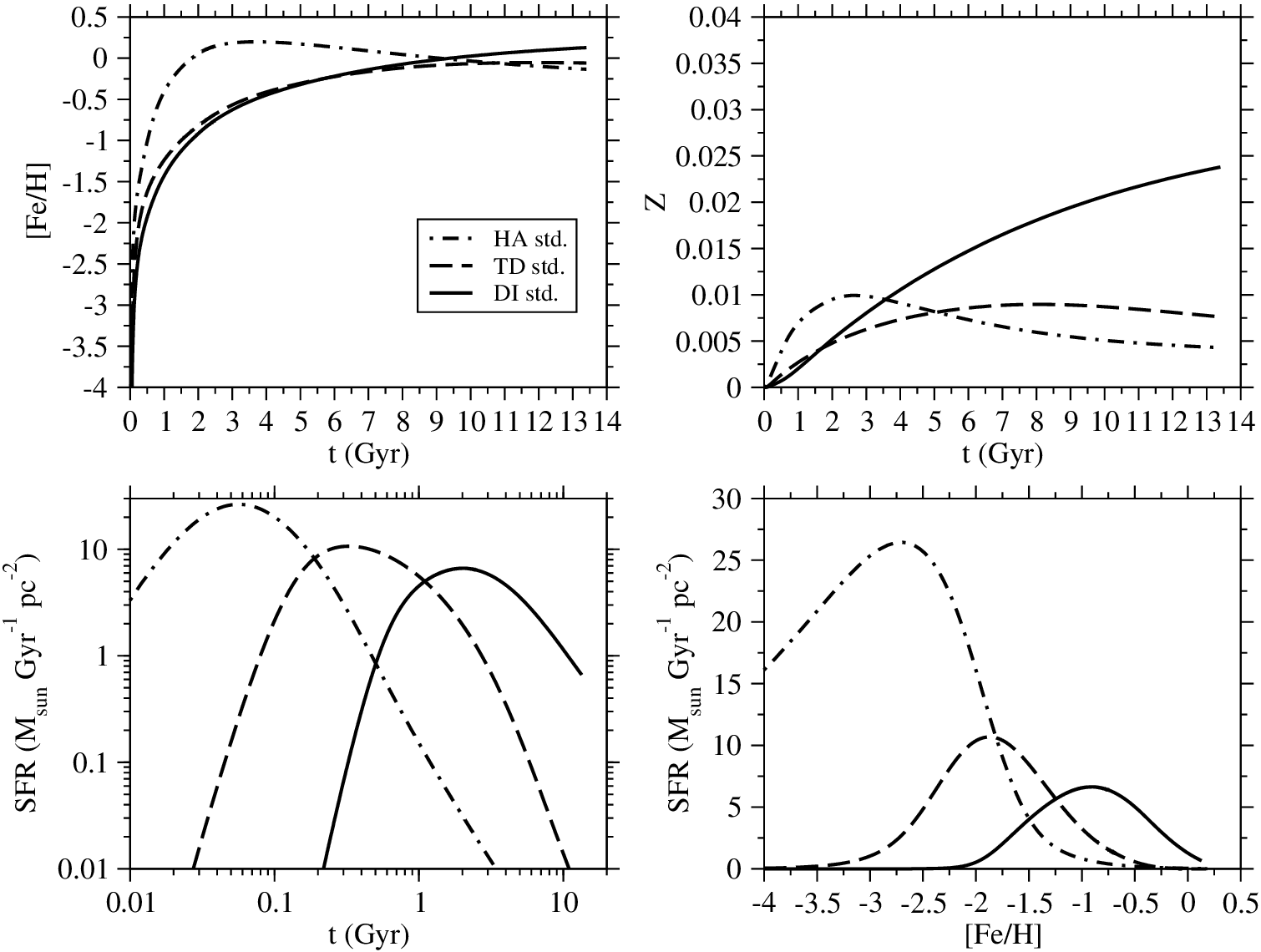}
\caption{Standard reference
model for our deterministic, constant coefficient, multi-population, 
multi-zone models.  Panels show the
evolution of three zones (halo, thick disk, thin disk):  clockwise
from upper left: metallicity evolutions ([Fe/H] and $Z$),  population
histogram for  stellar metallicity.  The SFR for fixed IMF 
provides a proxy for the number of stars at any instant 
for high mass stars with comparatively brief 
main sequence lifetimes.}\label{fig:all-std}  
\end{figure}

Matter exchanges between different zones and phases are codified by
 coefficients given by the typical timescale and efficiency of the
 related process.  In {\it our} standard framework we keep these
 coefficients at constant values during the evolution of the system,
 so any time dependence arises from the variation of the  mass
 fraction of the different phases. Some variations of the microphysics
 properties of the system over time might occur, which would affect the
 rates (Shore \& Ferrini \cite{sf95}).  The main rates  that drive the
 interactions among the different phases are the spontaneous formation,
 the cloud-cloud collisions terms, and the
 star-cloud interaction terms.   For  a detailed description
 of the standard version of the model -- in particular  the adopted
 nucleosynthesis prescriptions, SNI and SNII progenitors, initial mass
 function -- we refer the reader to Ferrini et al. (\cite{fmpp92}),
 Pardi \& Ferrini (\cite{par94}),  and  Ferrini et al. (\cite{fmpd94}),
 and Valle et al. (\cite{valle}).  For comparison with  subsequent
 simulations, we show the results of the \stdmdl\  in
 Fig.~\ref{fig:all-std}.  The coupling between zones is obtained
 through the infall parameters $f_H$ and $f_{TD}$.  The thick disk
 accumulates matter falling from the halo and in turn feeds the
 thin-disk zone.  These two coefficients are independent and kept
 constant in time; they are the only truly free parameters of the
 model. The timescale for the formation of the thick disk from the
 halo is assumed to be shorter than the accretion time for  formation
 for the thin disk during which it accretes  matter from the thick
 disk ($f_H = 0.1$, $f_{TD} = 6.5\times 10^{-3}$).

The model output includes the time dependent abundances of H, $^2$D,
$^3$He, $^4$He, $^{12}$C, $^{13}$C, $^{14}$N, $^{16}$O, $^{20}$Ne,
$^{24}$Mg, $^{28}$Si, $^{32}$S, $^{40}$Ca, $^{56}$Fe and $s$- and
$r$-process elements  (Travaglio et al. \cite{t99}).
The reactions for light isotopes (Li, Be, B)  production have recently
been inserted (Valle et al. \cite{valle}),  including the Galactic
cosmic ray spallation on interstellar CNO nuclei ({\em direct}
processes), spallation by CNO cosmic ray nuclei impinging on
interstellar $p$ and $^4$He ({\em inverse} processes), and
$\alpha$--$\alpha$ {\em fusion} reactions.  The model produces good
agreement with the linear trend of Be and B observed at low
metallicities without other modification to the model parameters or
equations.

This \stdmdl\ is {\it closed} in the sense that neither accretion into,  
nor outflow from, the system occur.  Any gas shed from the halo and
thick disk is assumed to accrete ultimately 
onto the thin disk and the mass of the 
galaxy stays constant.  The evolution computed using 
constant rate coefficients and constant IMF in 
time.  No instantaneous recycling is used, the metallicity evolution
is explicitly computed for a fixed initial mass function by including
stellar mass-dependent nucleosynthetic yields.  Turbulent mixing and 
advective transport are not included so the treatment is effectively
radially local and vertically stratified, in contrast to chemodynamical 
models, although it should be complementary (for a comprehensive survey 
of the various modeling schemes, see Hensler et al. \cite{hens03}).   
In the last panel of Fig.~\ref{fig:all-std}, and in subsequent 
figures, we use the SFR at any time to represent the number of 
stars formed at that instant with metallicity $Z(t)$.  This is, 
however, only a theoretical -- not observable -- representation for 
the metallicity distribution 
in any galactic zone ({\it cf.} Ferrini et al. \cite{fmpp92}).

\section{Stochastic Systems with Infall}
\subsection{Closed Systems}

There is overwhelming evidence that the Galaxy, and most extragalactic 
stellar systems, are subject to 
constant forcing by the combined effects of collisions, 
accretion, and stripping of gas and stars.  Our system has 
accreted at least a few 
dwarf galaxies in its lifetime -- for instance, Sgr and CMa -- and the high 
velocity clouds also seem to be infalling gas that isn't merely recycled 
from the corona or disk and may originate outside of the Galactic halo 
(Wakker et al. \cite{wak99}).  The detection of deuterium, D/H $\sim 10^{-6}$, 
in Galactic center molecular clouds is further evidence for an infall rate in 
the disk of about 1 M$_\odot$ yr$^{-1}$ (Lubowich 
et al. \cite{lubov00}).  But before 
examining the effects of any external mass accretion, we simulated  
closed systems assuming stochastic halo infall to the thick disk 
and thick disk infall to the thin disk as a baseline for the open cases.

\begin{figure}
\centering
\includegraphics[width=14.5cm]{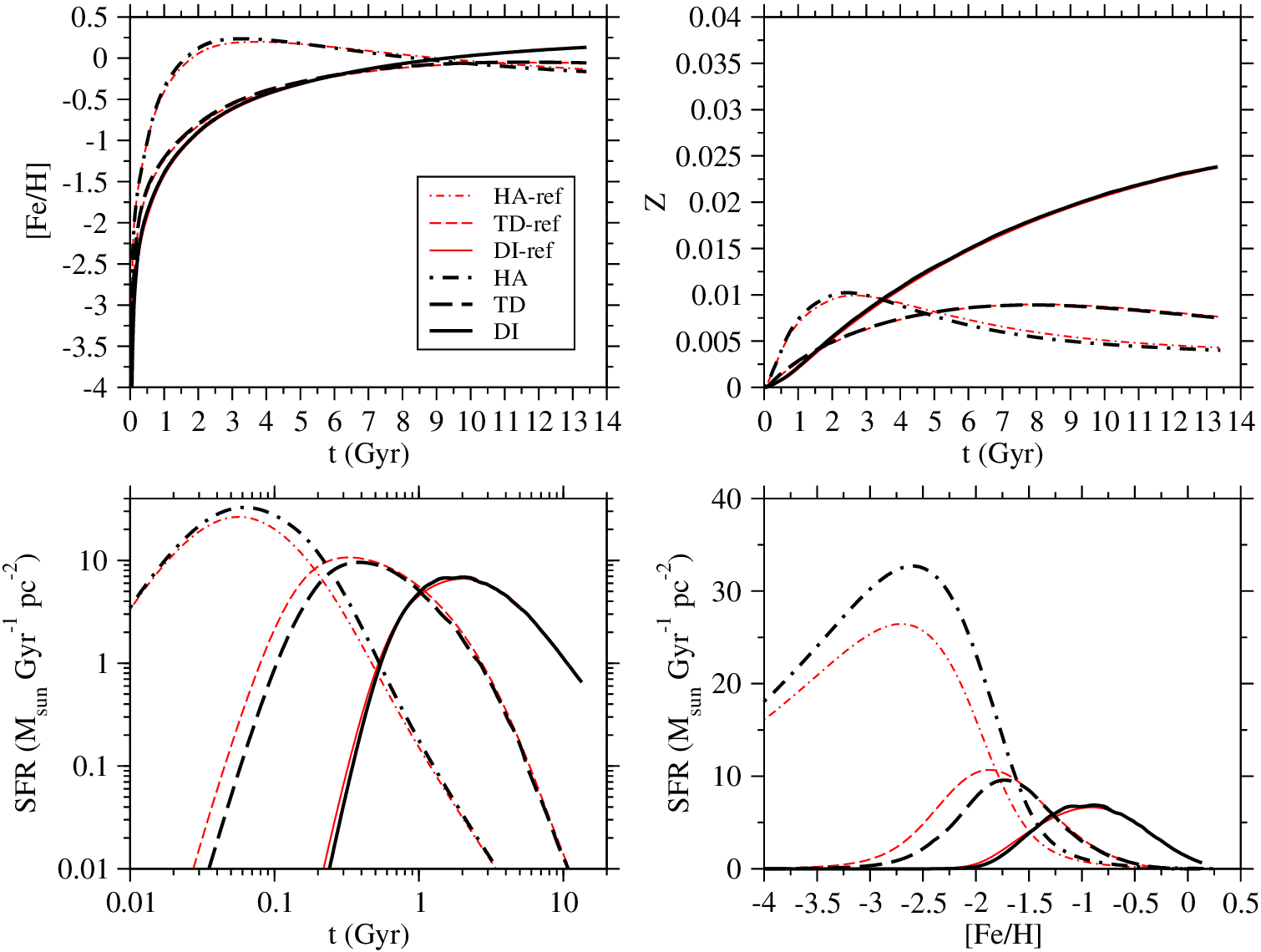}
\caption{Stochastic closed system: Sample simulation of random 
infall for a closed system (see text for
details).  The display is the same 
as Fig.~\ref{fig:all-std}. }\label{fig:all-stoc-clos}
\end{figure}

We modified the \stdmdl\   by adopting infall rates $(f_H,f_{TD})$
that fluctuate randomly in time  with the amplitudes being drawn from
a Gaussian  distribution with a freely chosen dispersion to mimic
shortlived accretion bursts that might occur for infall of halo or
thick disk clouds.  We  maintained a constant timestep for the models,
3 Myr, with an infall event  every 50 timesteps (150 Myr).  In each
event, which has impulsive rise and  exponential decline, the decay
time is specified as a parameter (in these models 100 Myr).
The mass of the total system was  kept constant.   A variety of models
were computed with  range of dispersions and amplitudes but these
produced  the same basic behavior so we will here report only the
results for  one simulation, shown in Fig.~\ref{fig:all-stoc-clos}.
This model  assumed that the mean  value for infall from each zone was
the same as the \stdmdl\ but with  $f_{H} : N(\mu = f_H, \sigma = 30\%
\; f_H )$,  and $f_{TD} : N(\mu = f_{TD}, \sigma = 30\% \; f_{TD})$
where $N(\mu,\sigma)$ is a normal distribution with mean $\mu$ and
dispersion $\sigma$.  Compared to the  \stdmdl, neither the
metallicity distribution nor the star forming history changed
significantly for the thin disk for closed system 
stochastic accretion.  The changes seen (Fig. 2) depend on the timing
of specific infall events.  For instance, enhanced infall early in the
evolution of the halo reduces the SFR while a negative fluctuation
enhances it; different realizations of a stochastic sample will
produce altered histories with the mean behavior mimicking the
\stdmdl.  Reverse
transfer between zones (e.g. blow-out, SNR, etc.) can also be included
in this prescription but the results are expected to be similar  to
those shown in Fig.~\ref{fig:all-stoc-clos} since the total mass of
the system does not  change.  The model assumptions are similar to
those of  Copi, Schramm, \& Turner (\cite{copi95}) and Copi
(\cite{copi97}) with one important difference:  because we use the
full set of model equations, random forcing also  causes the star
formation rate to vary in its own timescale in  response to the
accretion  fluctuations.  The combined effects of dilution and
resupply of gas for  continued star formation increases the dispersion
in each zone's elemental  abundance distribution but doesn't alter its
mean value.

\subsection{Open Systems}

To study the effects of external gas accretion, perhaps in the form of
high velocity clouds or intracluster gas, we performed a variety of
simulations of open systems.  In all the external diffuse gas, with
various assumed metallicities, was added directly to  the {\it thin}
disk.  The total mass of the three zones  increases by up to 44\% over
the simulation.   We should  emphasize here that our models assume, in
effect, a dark matter halo for the bulk of the gravitational
potential.  We do not need to include the dynamical reaction of the
three zones which  are passive tracers of the  chemical and star
forming histories.  Thus, although we have models for which  $M_{\rm
final}/M_{\rm initial}$ may be very far from unity, this is still a
small  change relative to the total galactic mass.  Both the
amplitudes and intervals  between successive events were randomized.
The times have been chosen to  examine the behavior of the model, not
for any particular reason related to  the epoch of formation of any
particular zone.  This contrasts with the  parametric study by
Chiappini et al. (\cite{chiappini97}) who by 
coincidence timed  their two infall
event  model to about the same instants as our simulations, about 1
and 3 Gyr,  but with the aim of mimicking the growth of the Galactic
thin and  thick disk.  These earlier models also adopted a
Schmidt-like power law  parametrization for the star formation rate,
in contrast to our multiphase  approach.

\begin{figure}
\centering
\includegraphics[width=6cm]{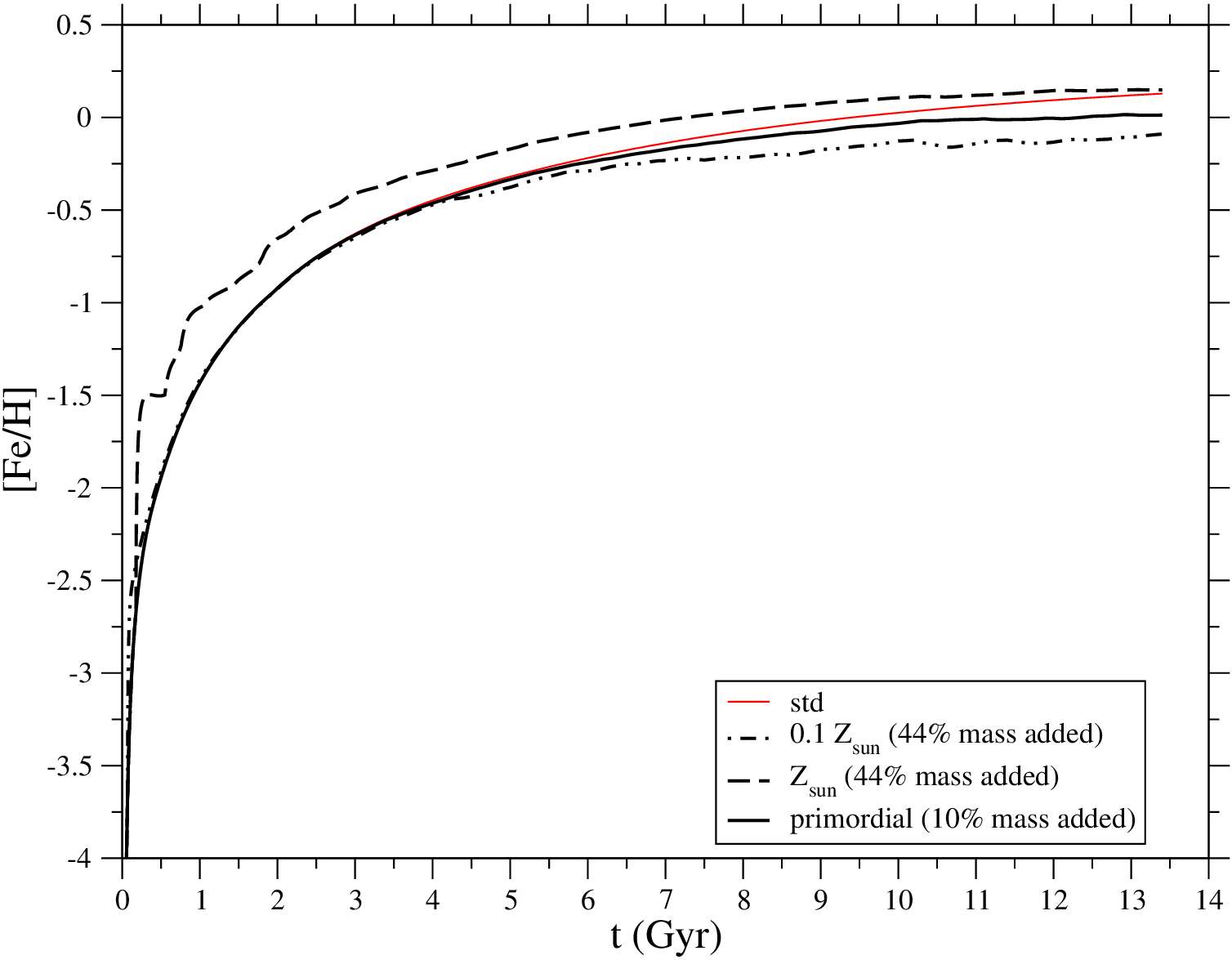}\hspace{5mm} 
\includegraphics[width=6cm]{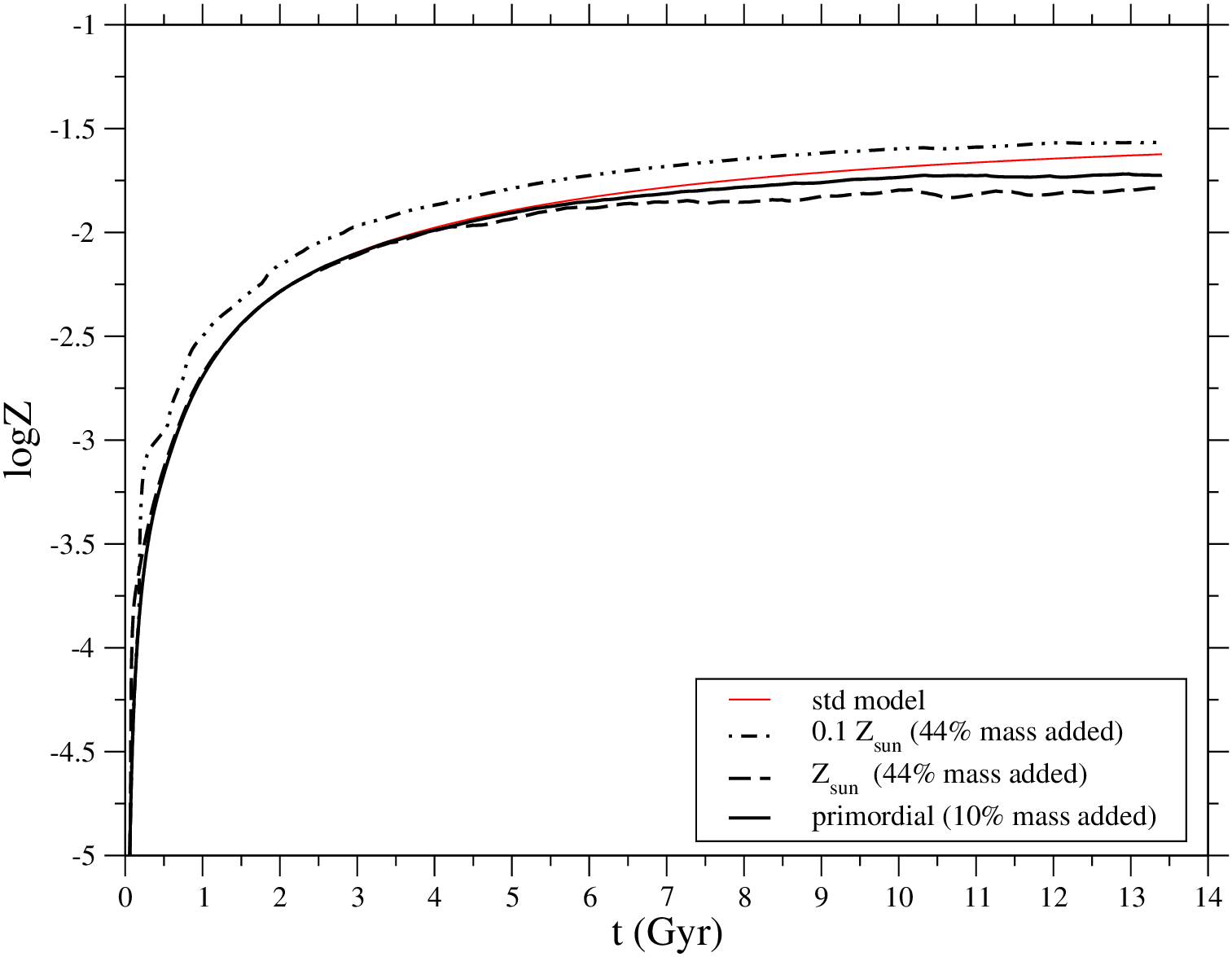}\vspace{5mm}\\
\includegraphics[width=6cm]{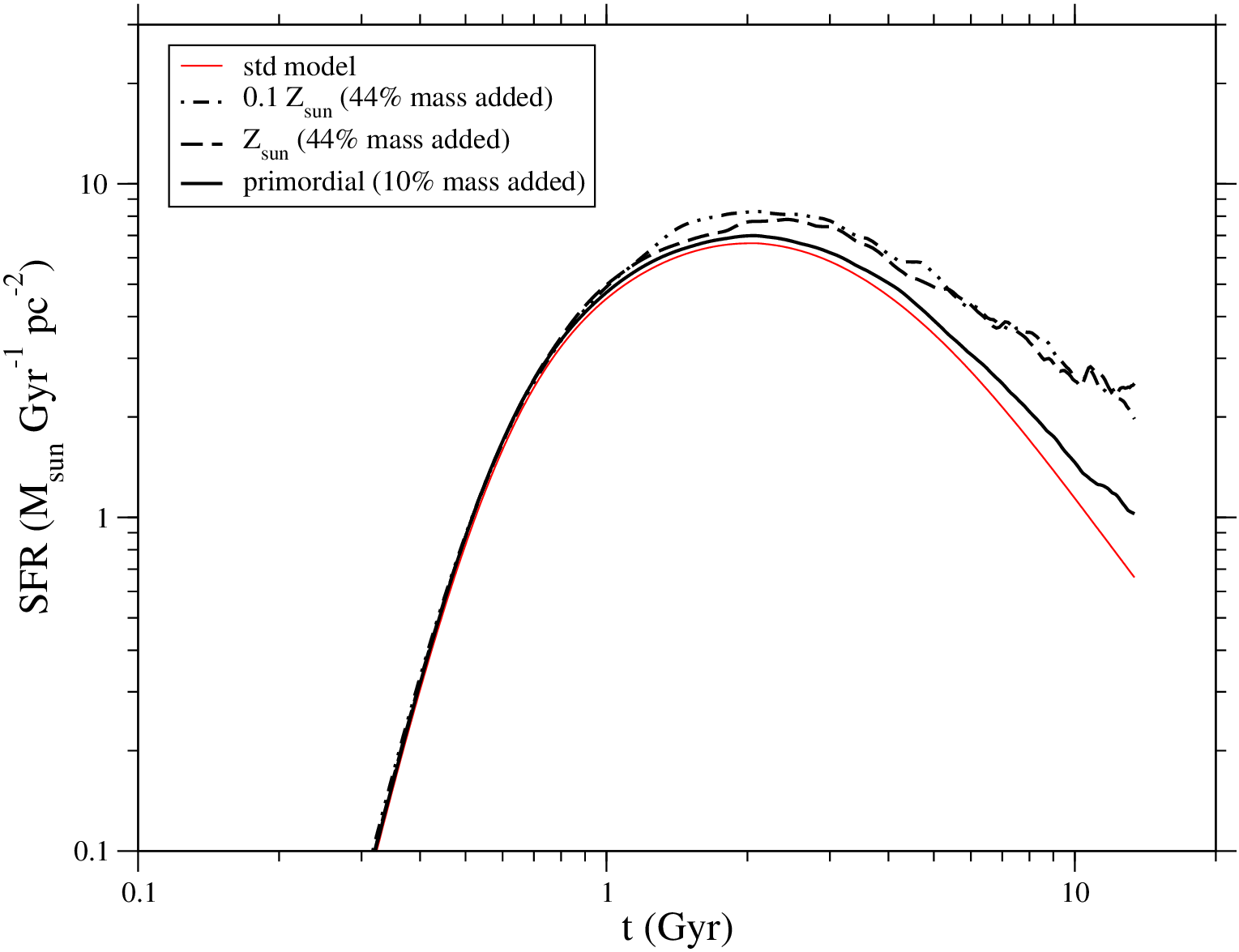}\hspace{5mm}
\includegraphics[width=6cm]{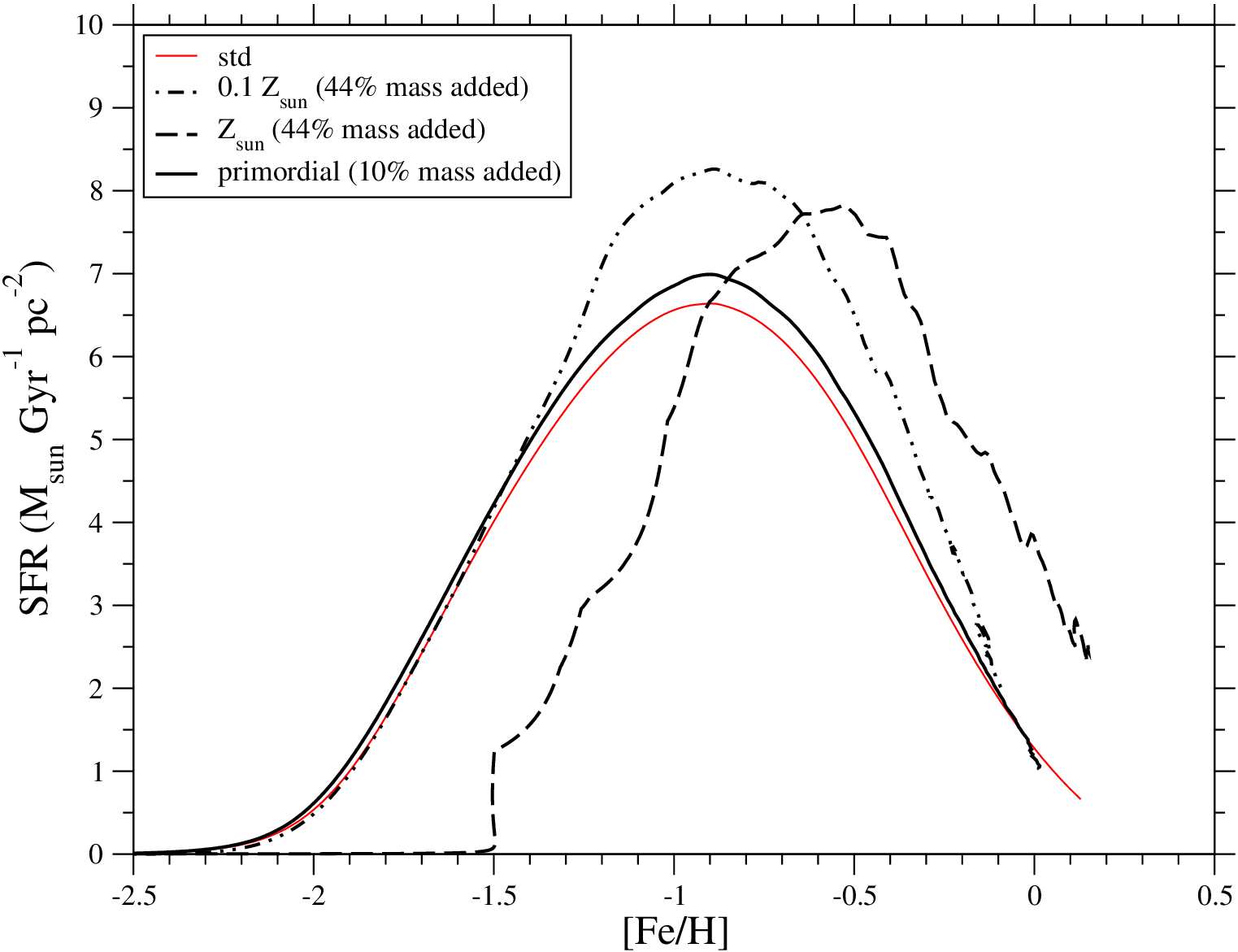}
\caption{Open stochastic simulation; 
in each case, the accretion is 
directly to the thin disk by an open system and only the thin disk
zone 
results are displayed compared to the standard model.  (Dash) 
Metal-rich  ($Z=Z_\odot$) stochastic accretion  (i.e. from the extragalactic 
environment) and the total mass increases by 44\%. Also shown 
are the cases for (dot-dash) 
low metallicity (Magellanic Stream simulation) with 44\% added total
mass, and (thick continuous)
primordial intergalactic gas with 10\% total added mass 
(see text for details).}\label{fig:open-accr-zsun}
\end{figure}

For infall of solar metallicity gas (Fig.~\ref{fig:open-accr-zsun}), the evolution is not drastically
changed  compared to the \stdmdl\ after the first few Gyr.  The final
thin disk  metallicity is higher throughout its history, reflecting
secondary  processing of the  already enriched gas.  The thin disk
metallicity distribution  shows a sharp lower cutoff at [Fe/H]=-1.5
and the metallicity and SFR stay consistently higher even at the end
of the calculation (by about 10\% and a factor of about 4,
respectively) relative to  the \stdmdl.  For the low metallicity
accretion case (Fig.~\ref{fig:open-accr-zsun}), we adopted 0.1 Z$_\odot$  for gas with the metallicity
of the Magellanic Clouds (to mimic gas  from the Magellanic Stream).
A similar metallicity has been proposed by  Wakker et al. (1999) for a
high velocity cloud.  The final abundances are reduced relative to
the \stdmdl\ but only by about 30\% while the SFR increases by about
the same  factor as the high-Z case since this depends only on the
amount of gas added  to the system.  But as a consequence, the [Fe/H]
histogram is essentially  unchanged from the \stdmdl\ although it
shows a high end cutoff at  [Fe/H]$\approx$-0.1 reflecting the
dilution by the accreted gas.

\begin{figure}
\centering
\includegraphics[width=11cm]{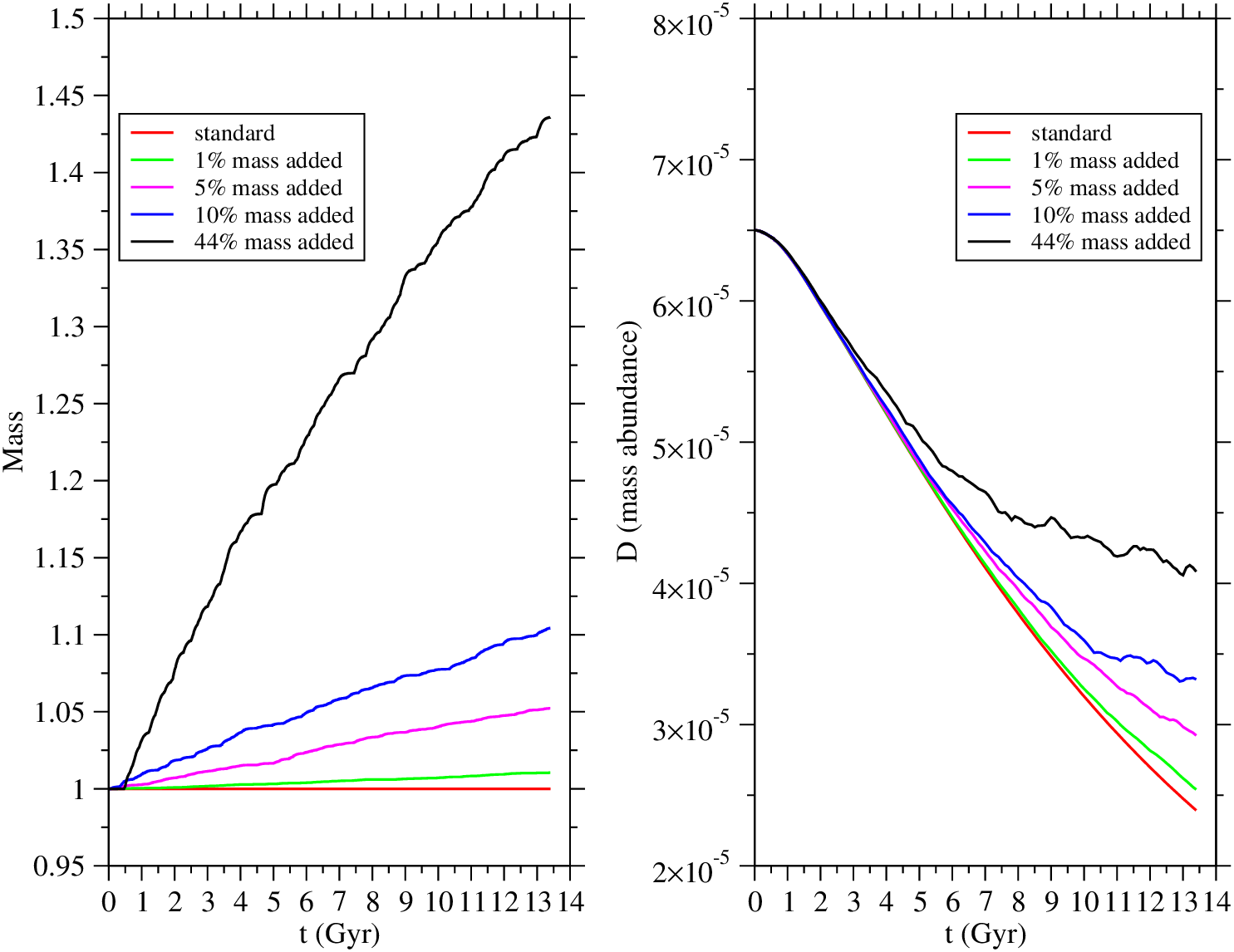}
\caption{$^2$D evolution for the open system described in 
Fig.~\ref{fig:open-accr-zsun} in the case of primordial material accretion. The
  temporal development of total galactic mass (a) and $^2$D 
thin disk abundance (by
mass)  (b) for a set of open stochastic accretion models with 
primordial gas accreted.
From the top to the bottom the curves refer to: 44\%, 10\%, 5\%, 1\% 
and no mass added.}\label{fig:mass-e-deu}
\end{figure}

Finally, to simulate primordial gas accretion, we show in 
Fig.~\ref{fig:open-accr-zsun} the result for stochastic infall for gas with 
standard Big Bang Nucleosynthesis (SBBN) abundances ($X_0=0.758, \; Y_0=0.242, \;
Z_0=0, \; ^{2}$D$=6.5\times 10^{-5}$).  The accretion, which again passes
directly to the thin disk, increases the system mass by 10\%.  At this 
low accretion rate, the thin disk SFR is unchanged.   
The metallicity evolution curve (top right)  illustrates the relatively 
weak dilution effects
that are offset by continuing star formation driven by the infall 
directly to the thin disk; the final metallicity is reduced by 
about 20\% relative to the \stdmdl\ while the histogram for [Fe/H] is 
unchanged.  For various infall rates the mass increase 
for the system is shown in 
Fig.~\ref{fig:mass-e-deu}{\it a} while the deuterium abundance 
evolution is shown 
in Fig.~\ref{fig:mass-e-deu}{\it b}.  

The early evolution of the $^2$D abundance  is dominated by astration.  
All models follow the same curve during the period of intense halo
and thick disk star formation which depletes the internal gas of deuterium. 
The subsequent evolution for the thin
disk is dominated by accretion and the competition between the 
$^2$D-depleted material shed from the halo and thick disk stars and the 
external accretion.  This reduces the final $^2$D abundance below SBBN.  
If the dynamical
mixing  timescale in the Galaxy is longer than the accretion rate,
some residual fluctuations can occur in the interstellar $^2$D 
abundance at the current epoch.

\begin{figure}
\centering
\includegraphics[width=8cm]{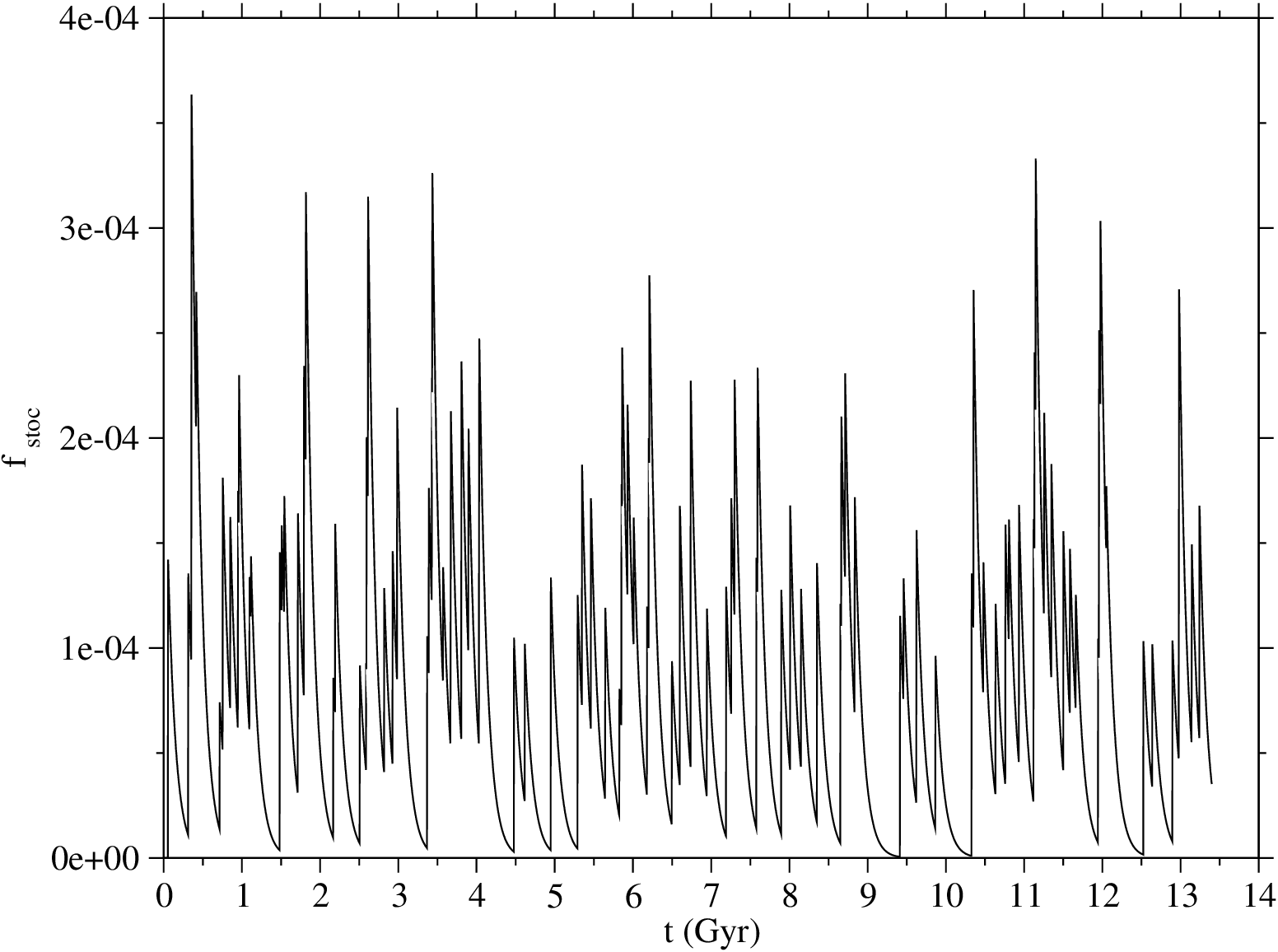}\hspace{10mm}
\includegraphics[width=8cm,height=6cm]{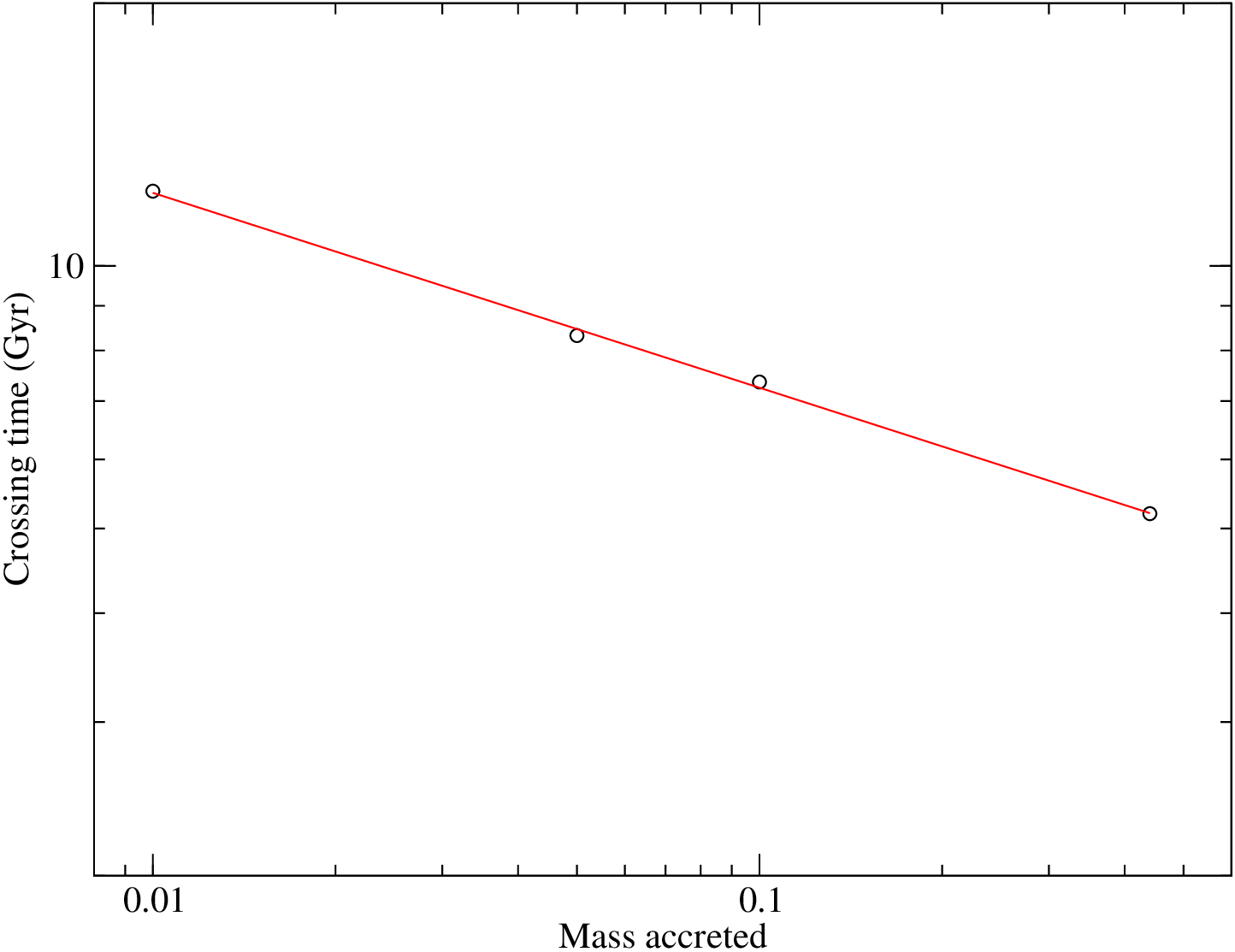}
\caption{(a) Single realization
illustrating the time history for the events for the model shown in
Fig.~\ref{fig:open-accr-zsun} in the case of primordial accretion.  (b) Dependence of the crossing 
time for galactic deuterium 
on fractional accreted mass (see Fig.~\ref{fig:mass-e-deu}(a)) for the galaxy for 
$\Delta$ $^2$D/$^2$D(std)=0.05. As explained in the text, this is a
way to quantify the possible abundance 
dispersion in a mixed galactic ISM and to
illustrate the competition between stellar processing and
environmental resupply.}\label{fig:cross-times}
\end{figure}

Fig.~\ref{fig:cross-times}{\it a} shows a sample realization 
of a single open stochastic model for the primordial case 
for low mass accretion.  Each infall 
event, of which there are 100 in this model, 
is assumed to have a short decay time, 
$\tau=$100 Myr.  In Fig.~\ref{fig:cross-times}{\it b} we show one 
way of quantifying the 
infall models by plotting the time at which the $^2$D abundance 
departs by a specified fraction from the predictions of the standard closed model as a function of the fractional accreted mass.
Fluctuations in D/H of order 5\% are 
compatible with ISM abundance studies and 
indicate the influence of infall relative to
mixing within the disk.  If the disk mixing time is 
less than the {\it crossing 
time} in Fig.~\ref{fig:cross-times}{\it b}, 
the process is likely to wash out local fluctuations.

\newpage

\section{Collisions and Stripping}
  
\begin{figure}
\centering
\includegraphics[width=14.5cm]{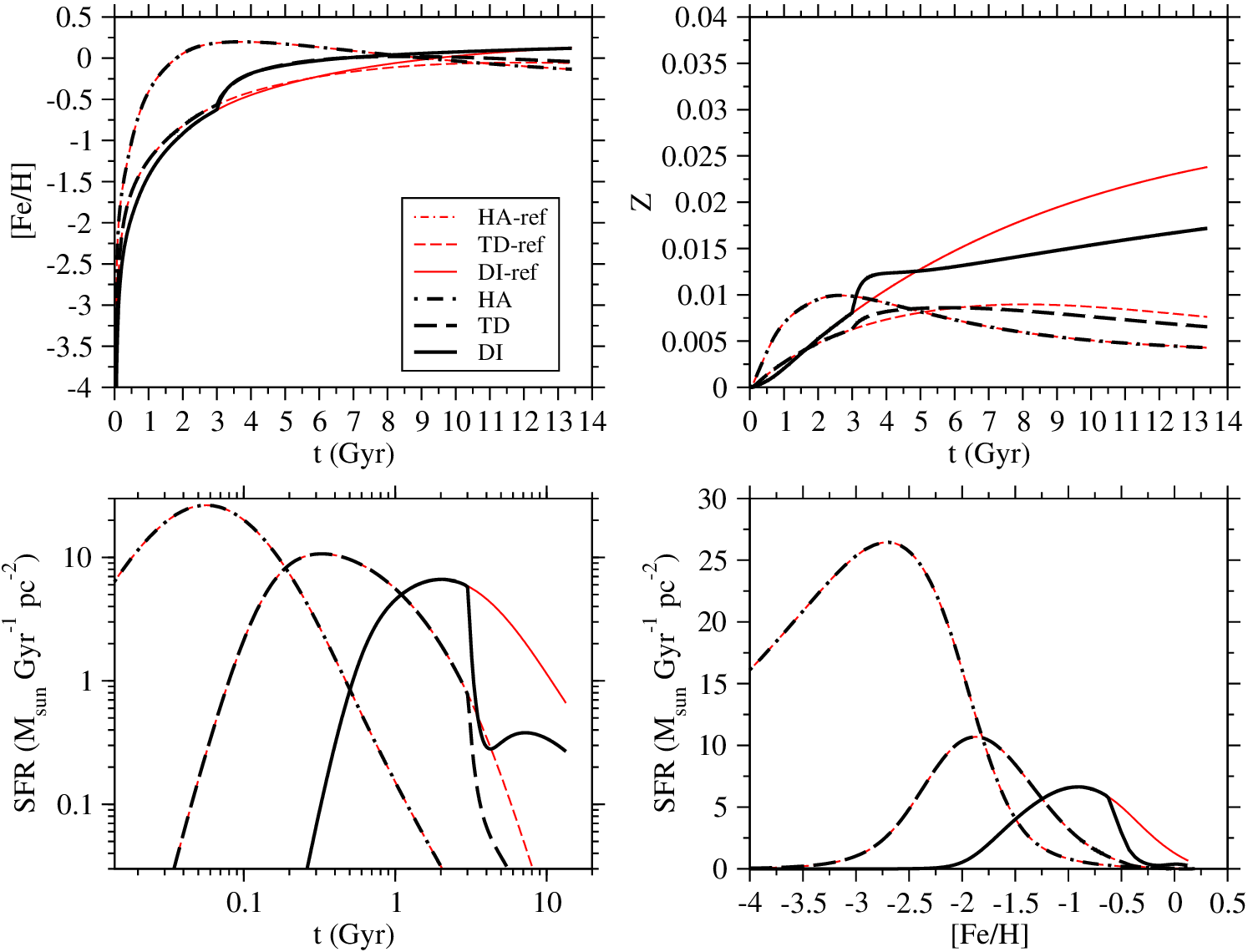}
\caption{Collision simulation starting at $t_{start}=3$ Gyr 
lasting for $\Delta t_{coll}= 50$ Myr (see text)}\label{fig:coll-50myr-3gyr}
\end{figure}

For any cluster member, collisions 
are always possible (Mihos \cite{mihos04}).  While 
we cannot model the different types in detail ({\it e.g.} 
major vs. minor mergers, for tidal disruption in the cluster potential), we 
can study the phase response to a schematic forcing.  
Collisions were simulated as impulsive events 
starting at some time $t_{start}$ and 
lasting for $\Delta t_{coll}$ during which
time the diffuse gas fraction is set continuously to some reduced
value, including complete removal.    
For the stripping simulations, the clouds 
were not removed during the events since studies of cluster 
galaxies imply that only the diffuse gas is removed ({\it e.g.} Moore
2004).  To test the effect of cloud removal during collisions, we have 
also performed simulations with various efficiencies of gas removal.  
During real collisions, galaxies appear to experience episodes of 
stimulated enhanced star formation.  We do not obtain this in our
models nor do we include its effect on the metallicity evolution by
arbitrary additional assumptions.  By keeping the 
rates that govern star-cloud conversions constant, we ignore (and therefore 
underestimate) possible triggered bursts, but our approach can
accommodate such scenarios when appropriate rates are physically 
calculated outside the model. Cloud destruction 
depletes that phase and, for this reason, the SFR drops only after the 
removal of the diffuse gas on the cloud destruction timescale. 
At the end of the collision interval, gas is
resupplied only internally, {\it i.e.} from mass shed by stars
that have evolved within the various zones -- without
additional mass loss from the system -- on a timescale determined 
by stellar evolution and the assumed IMF.  This approximates 
a fast, shortlived analog of stripping that lasts 
only a short time after which gas removal stops
abruptly.   In the cases shown here we used durations of 50 and
130 Myr started at two different times: around 
the peak of the \stdmdl thin disk star formation (3 Gyr),  
and before the peak in the thin disk \stdmdl
but after the peaks of the halo  and thick disk (1 Gyr).  Different 
efficiencies were assumed for the fraction of gas removed in
a single event (the stripping efficiency) and we used a variety 
of time profiles.  We present only the results for single or 
double collisions in Figs. ~\ref{fig:coll-50myr-3gyr} and 
~\ref{fig:coll-130myr-1e3gyr}.  
Any plausible combination of waiting times and amplitudes 
for any of these events produces a superposition of the
outcomes of the individual collisions.  

\begin{figure}
\centering \includegraphics[width=6.5cm]{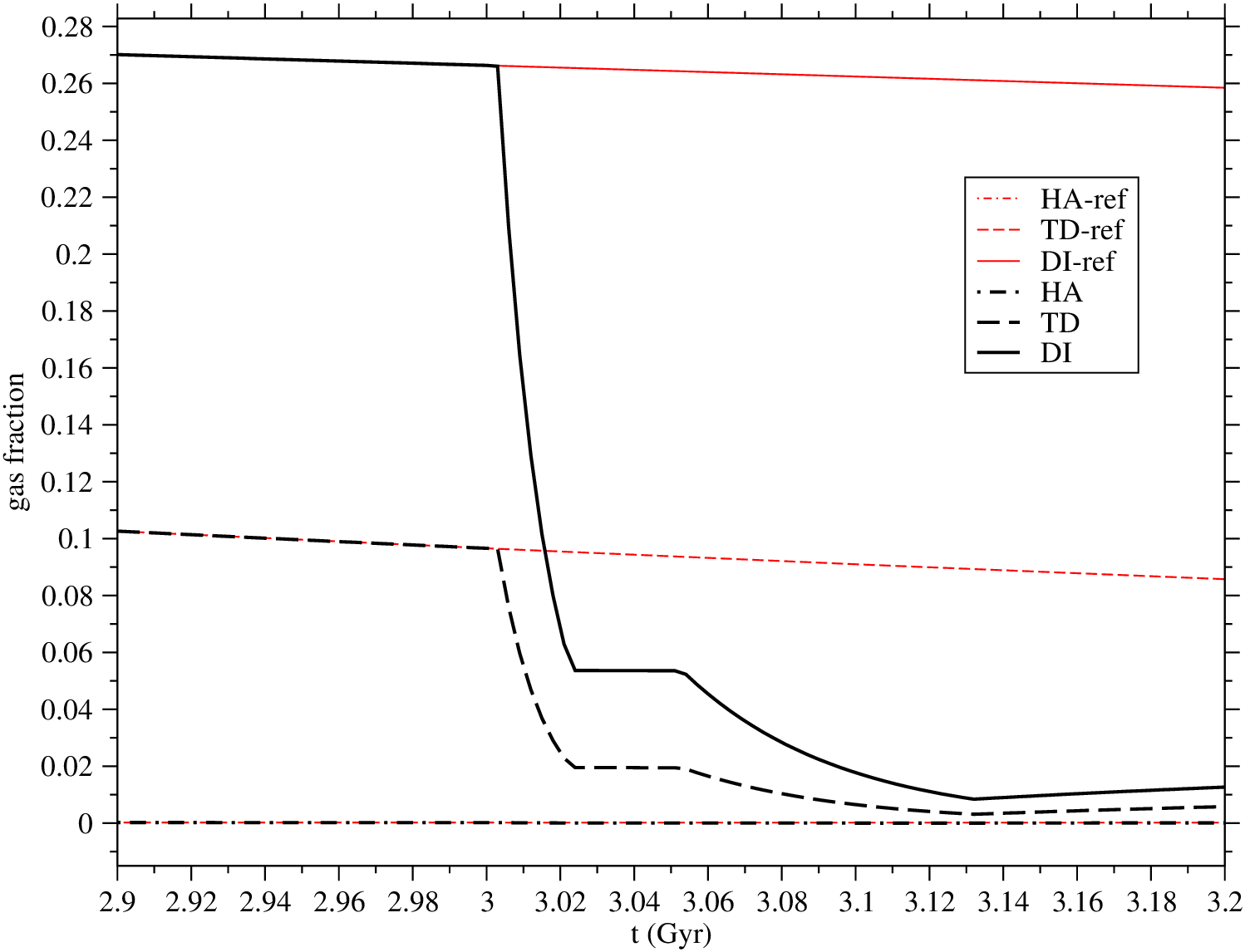}
\caption{Time histories for the mass of the galaxy in 
collision simulations using the mass loss profile from 
Quilis et al. (\cite{quilis00}) for $t_{start}=3$ Gyr 
and $\Delta t_{coll}= 130$ Myr.}\label{fig:gas-coll}
\end{figure}

\begin{figure}
\centering
\includegraphics[width=8cm]{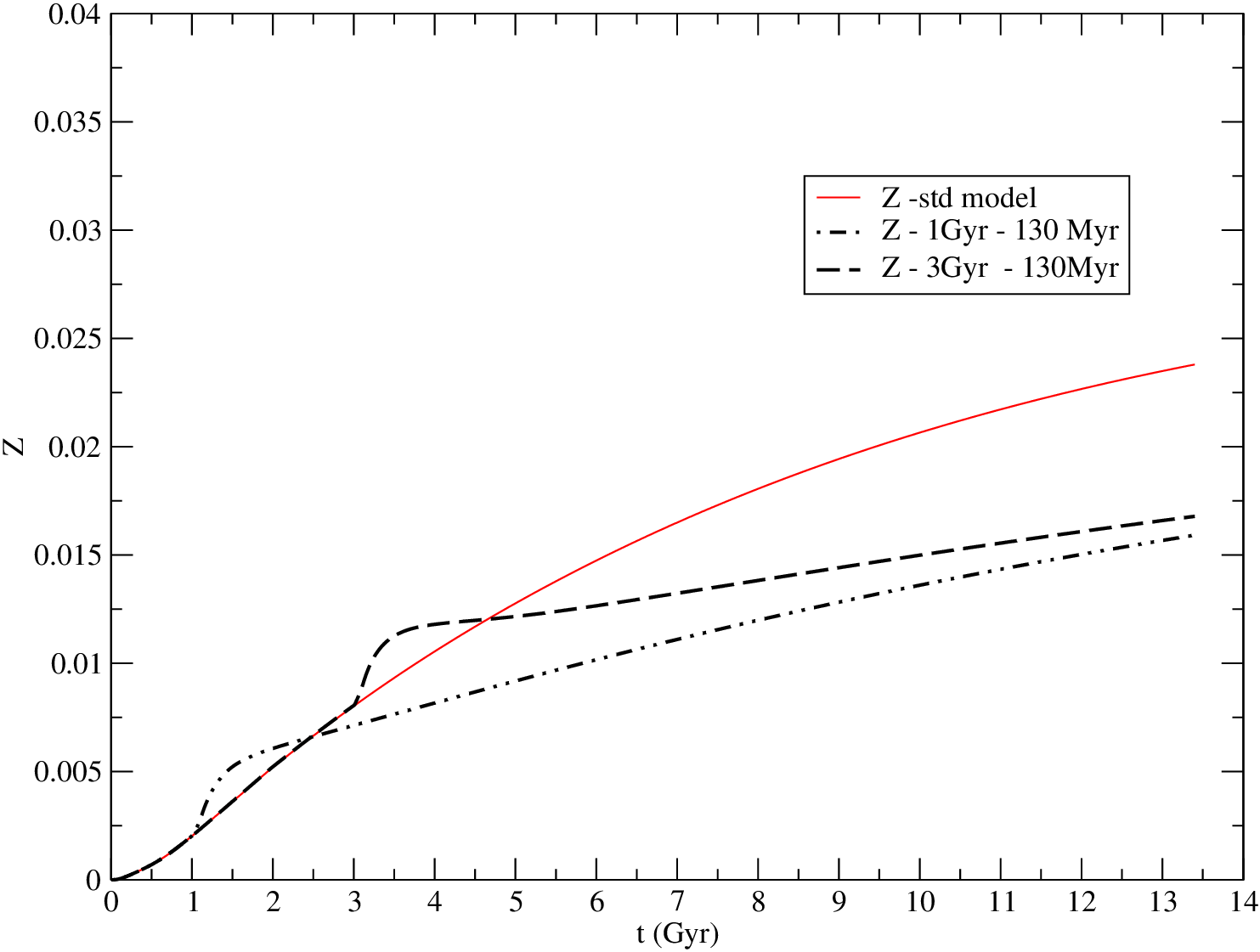}\hspace{10mm}
\includegraphics[width=8cm]{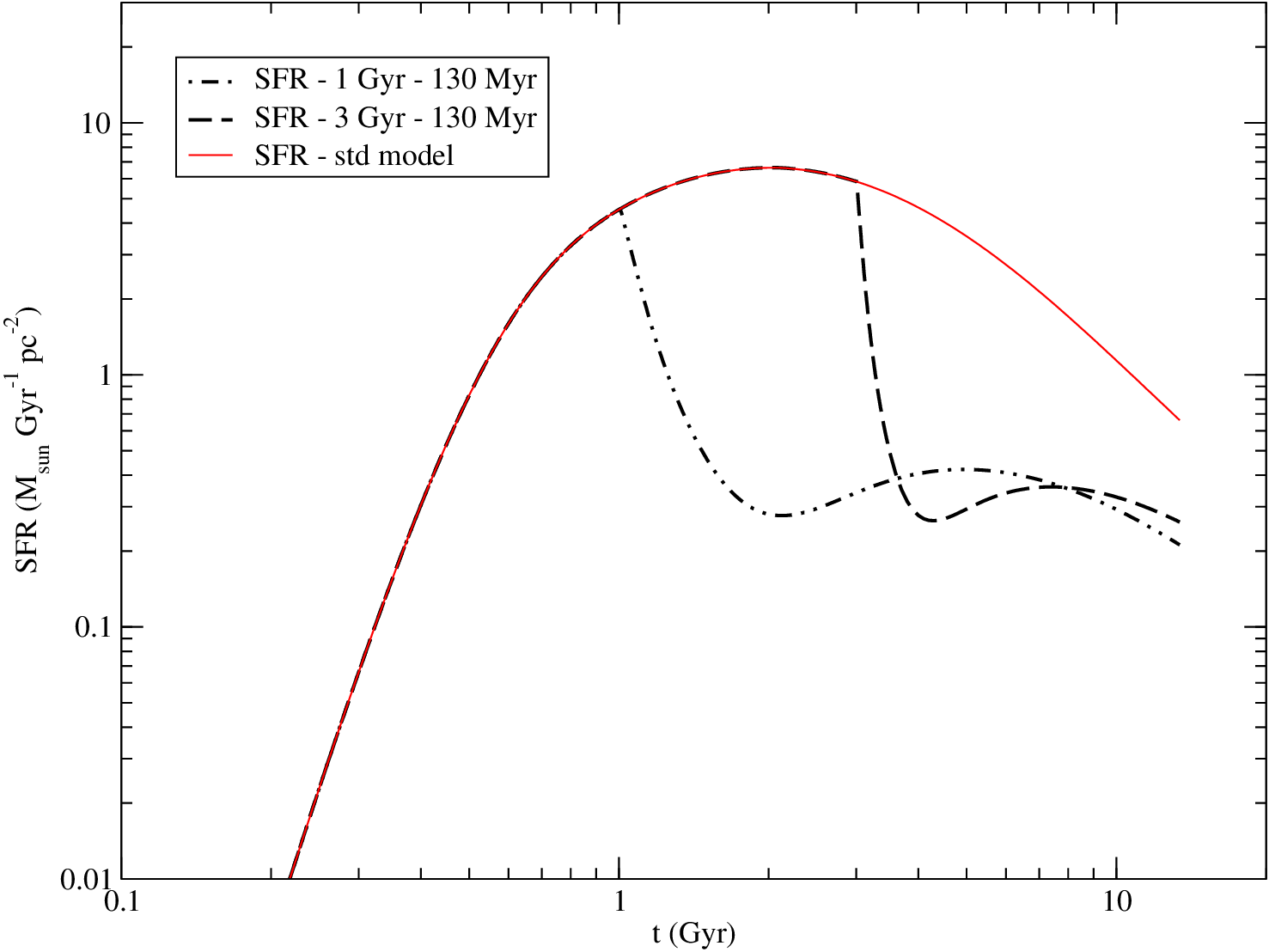}
\caption{Collision simulations with $t_{start}=1$ Gyr, $t_{start}=3$
Gyr and  $\Delta t_{coll}=130$ Myr. Residual gas 3\% after $\Delta
t_{coll}$.  Mass loss profiles were as in Quilis et al.
(\cite{quilis00}). Note: 
In this and all subsequent figures we display only the thin disk
evolution.}\label{fig:coll-130myr-1e3gyr}
\end{figure}

To attempt to include a dynamically modeled mass loss history within the 
stripping during a collision, we adopted the 
numerical results of Quilis, Moore, \& Bower (\cite{quilis00}), shown for 
reference in Fig.~\ref{fig:gas-coll}. This stripping history governs 
all diffuse gas within the galaxy including that returned by stellar mass
loss. We find a general behavior of the systems. The 
models are quite insensitive to 
the detailed  {\it history} of the event but depend strongly on the 
efficiency of the gas removal in the initial 
stage of the collision and on the {\it timing} of the 
collision, since the removal of gas throttles the star formation and 
produces a hiatus in the metallicity production.  
Since we compute the SFR directly, 
we see that its sharp drop halts metallicity production for some time, until 
stellar evolution resupplies disk gas and heavy elements.  The recovery 
is not as complete as first found by Comins \& Shore
(\cite{com90}). We note that in these older models, an 
improper normalization to the total mass was applied.  Consequently the
efficiency of replenishment of the gas by stellar death was
overestimated. We now find that the decrease in the 
active mass means the SFR after the collision never reaches 
its previous levels or those of the \stdmdl\ and the system remains 
permanently metal poor.  This may reproduce the metallicity histories 
of dwarf galaxies in clusters that undergo very early tidal interactions 
while still forming stars 
(see Fig.~\ref{fig:coll-50myr-3gyr} and Fig.~\ref{fig:coll-130myr-1e3gyr}).  

During a collision, the metallicity initially rises by about 20\% over 
a very short time, about $\Delta t_{coll}$.  Thereafter, 
the reduced SFR produces 
a nearly flat, reduced metal abundance and an upper cutoff to the [Fe/H] 
vs. SFR histogram.  To test the sensitivity of the results to the duration 
and timing of
the collision, we varied the two parameters while keeping the Quilis et al. 
(\cite{quilis00}) profile for the mass loss.  The results are shown 
in Fig.~\ref{fig:coll-130myr-1e3gyr}.
   
For collisions with $t_{start}=3$ Gyr, the upper cutoff 
is at about [Fe/H]=-0.5.  This is drastically altered for the earlier 
$t_{start}$ models for which the thin disk never effectively forms.  The 
histogram is significantly reduced with a peak at around [Fe/H]=-1.5 and 
the thick disk also developed a reduced metallicity, a cutoff is found 
for [Fe/H]$\approx$-1.  The final $Z$ for the thin disk is reduced by 
about 40\% relative to the \stdmdl.  

\begin{figure}
\vspace{2mm}
\centering
\includegraphics[width=8cm]{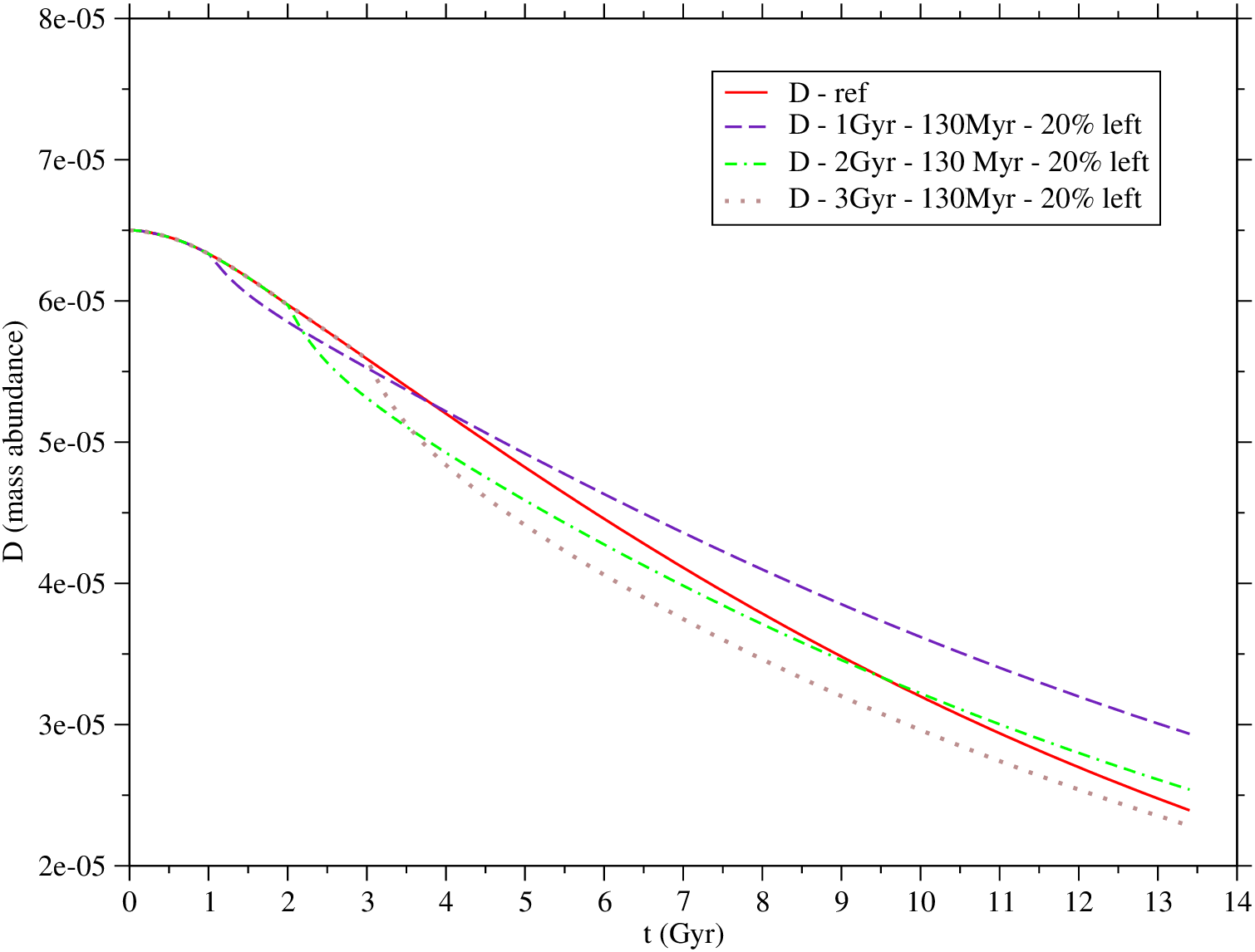}\hspace{10mm}
\includegraphics[width=8cm]{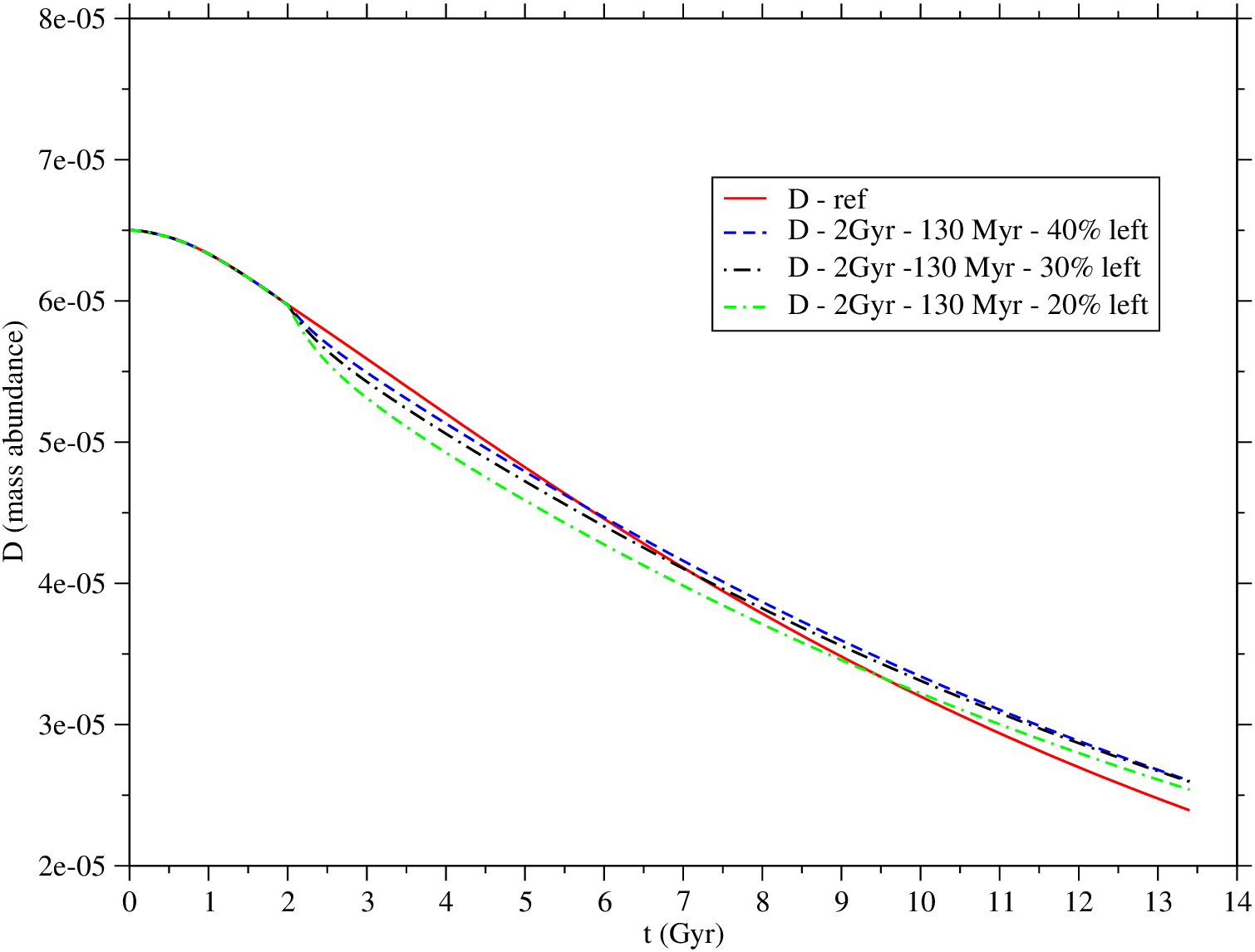}
\caption{(a) Thin disk D abundances in collision depending on starting time
(b) D depending on fraction of gas removed}
\label{fig:deu-tstart}
\end{figure}

\begin{figure}
\centering
\includegraphics[width=7.5cm,height=5.55cm]{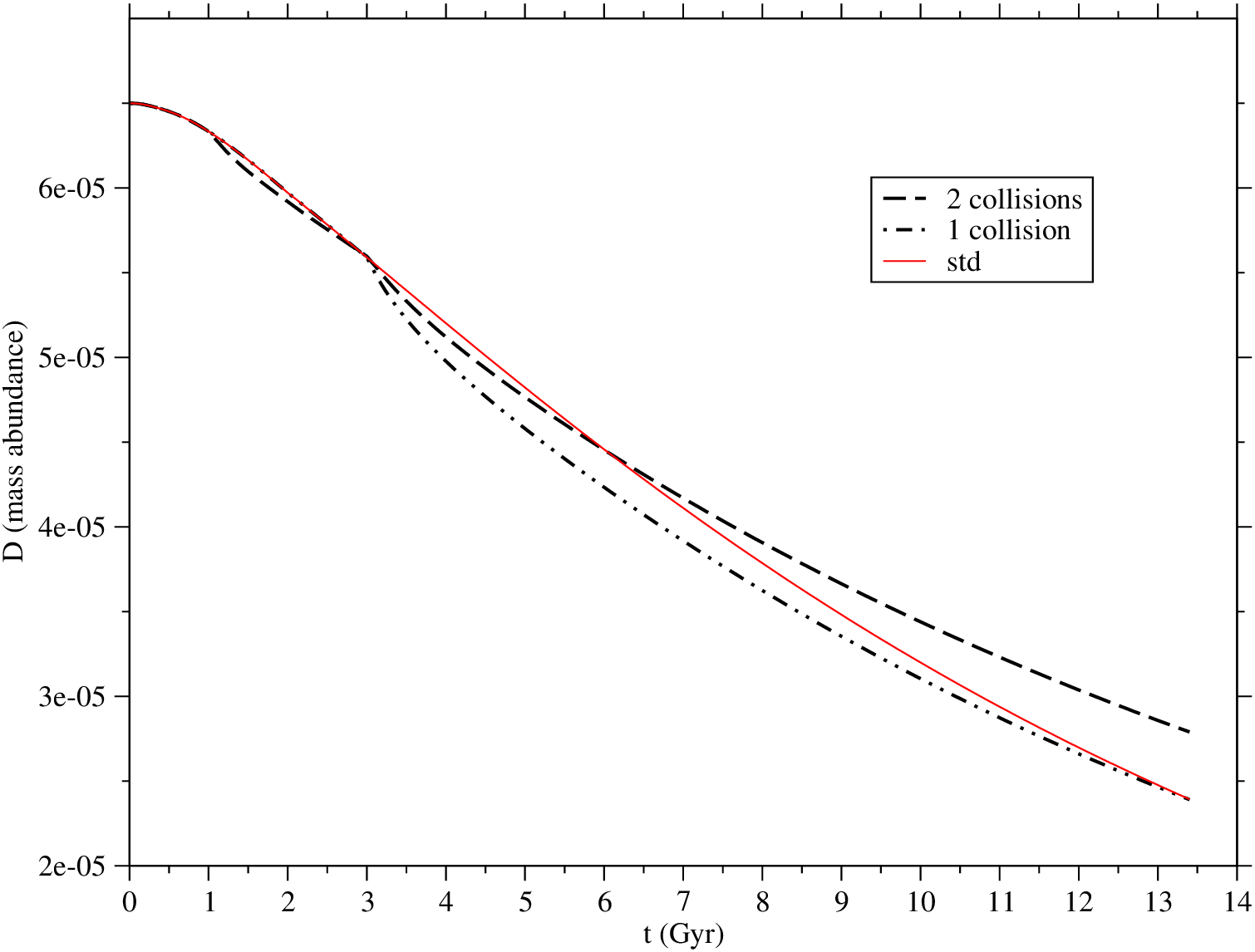}\hspace{10mm} 
\includegraphics[width=8cm]{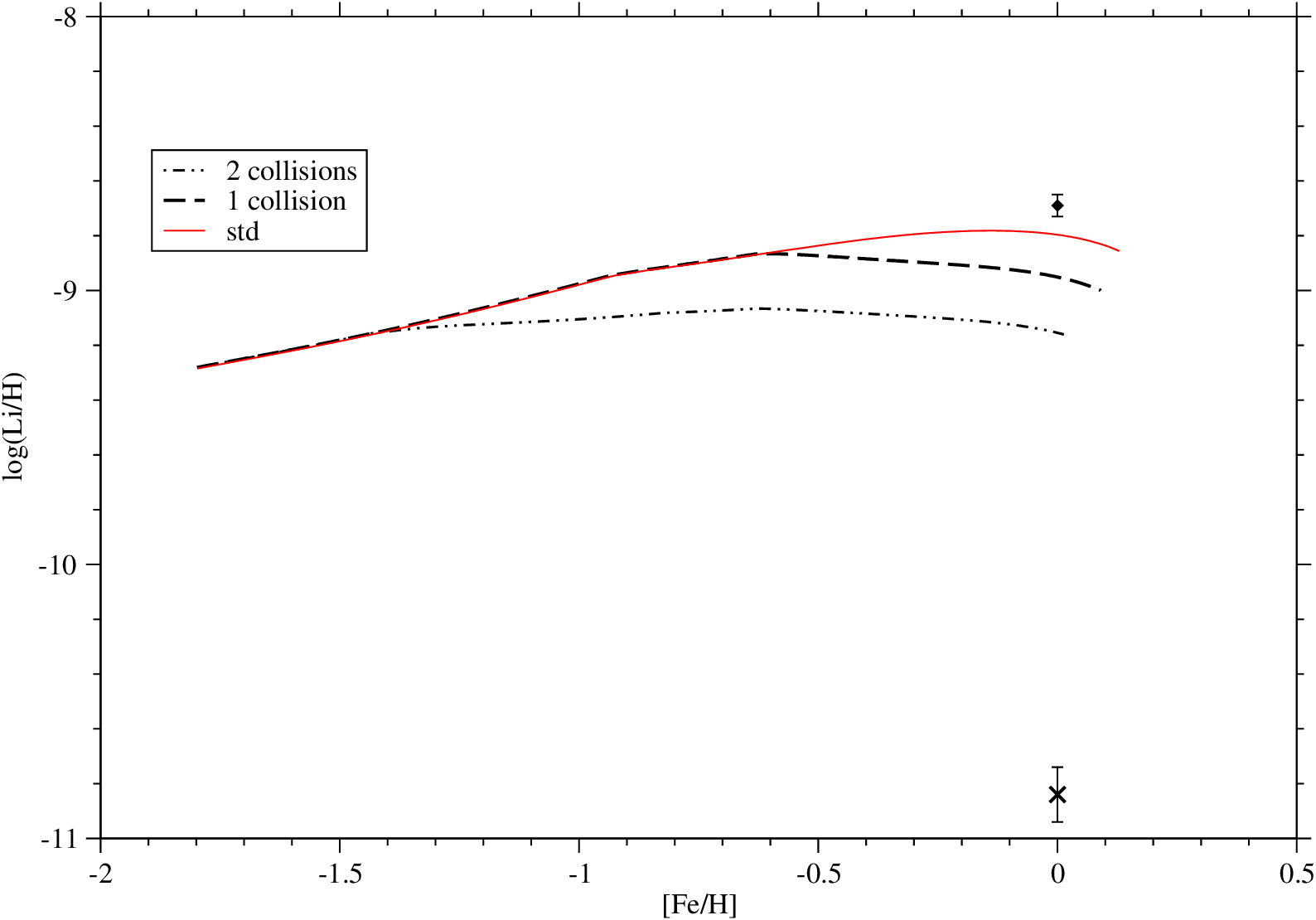}\vspace{3mm}\\
\includegraphics[width=8cm]{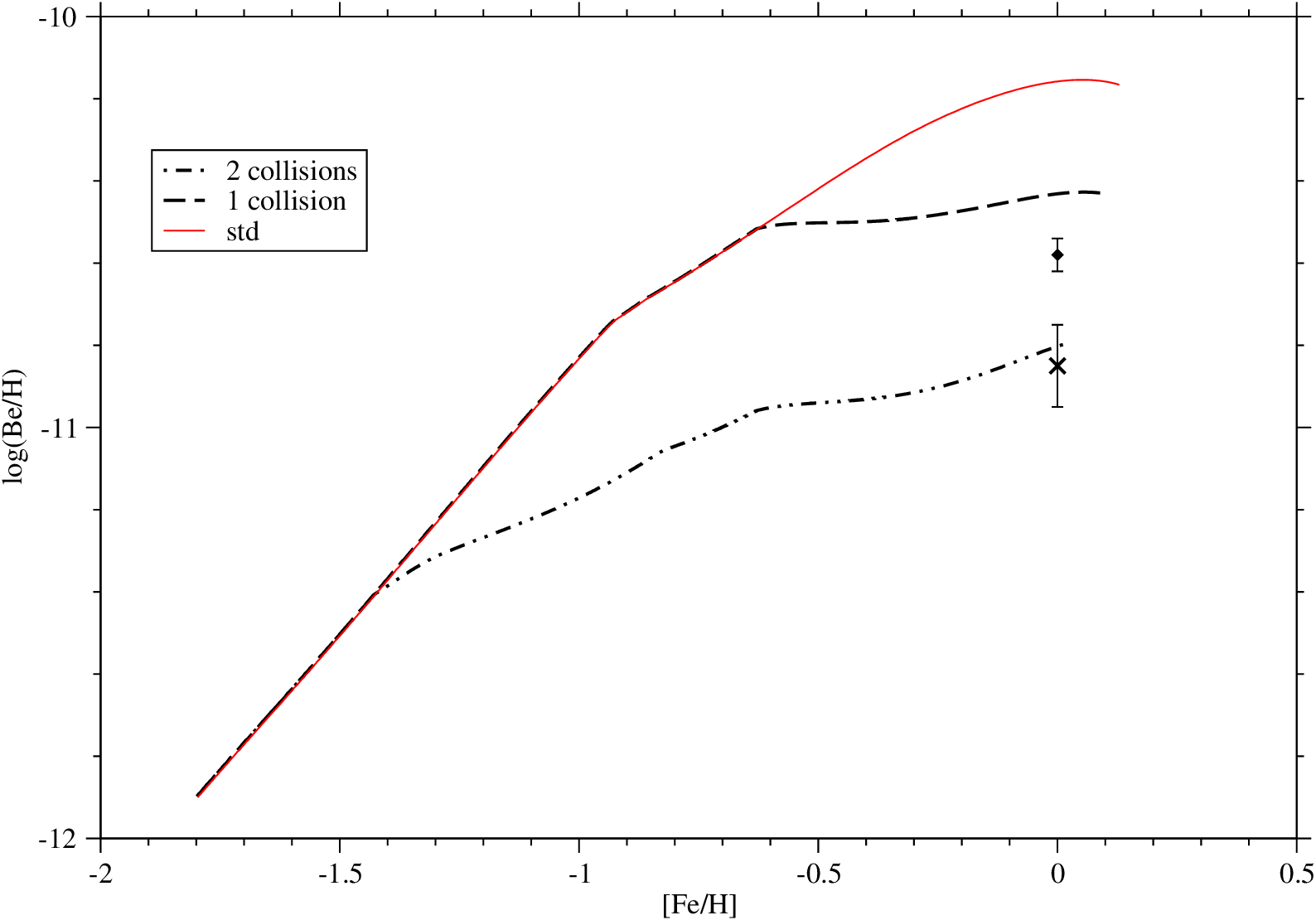}\hspace{5mm}
\includegraphics[width=8cm]{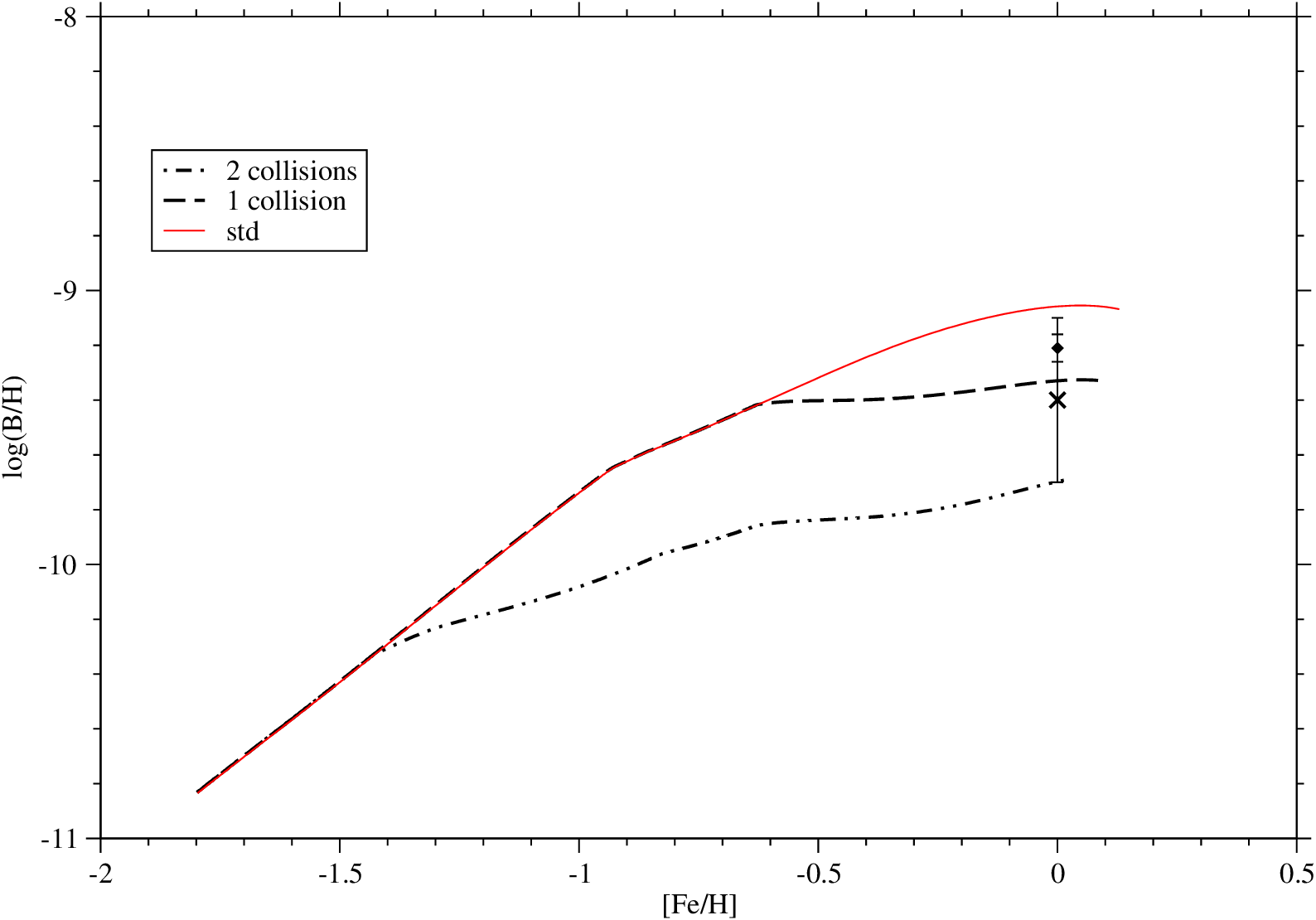}
\caption{Comparative evolution of the light elements in two
collisional models for the thin disk zone (see text):  
a single collision as in Fig. 8
with $t_{start}$=3 Gyr and $\Delta t_{coll}$=130 Myr (dashed line) 
and the case of
two collisions with $t_{start}$=1 and 3 Gyr but the same duration 
(dot-dashed line).
The panels shows, in order from the top, $^2$D, total Li, $^9$Be,
and total B.  For Li, $^9$Be, B, the meteoritic and the solar photospheric abundances (from Grevesse \& Sauval \cite{greve98}) are indicated by a diamond and a cross, respectively. In all cases, the diffuse gas mass remaining was
30\% of its initial value after each collision.}\label{fig:all-light}
\end{figure}

As a final result from the collision simulations, we show in
Fig.~\ref{fig:deu-tstart} the deuterium evolution of the models.  For
this isotope, the curves are almost identical to the infall case for
relatively low metallicity gas. This  can be understood from the SFR
profile. If star formation is suppressed  so is astration ({\it e.g.}
Vidal-Madjar \cite{Vidal02}, Lemoine et al. \cite{Lemoine99}, H\'ebard
et al. \cite{hebard02}).  Since we found that the subsequent $^2$D
evolution  depends on the infall rate, the final abundance is set
mainly by the  amount of gas remaining in the system and resupplied
from the halo and thick disk.  Reducing the star formation for the
subsequent  history elevates the final abundances relative to the
\stdmdl\ (Lubowich  et al. \cite{lubov00} find  this for the Galactic
center region, for instance).   For multiple collisions, the results
for which are shown  in Fig.~\ref{fig:all-light} {\it b}, the $^2$D is
nearly  indistinguishable from the \stdmdl.

\newpage
\newpage

The effects  on the evolution of the light elements is, however,
striking since these  are produced by the continuing star formation.
While in  the \stdmdl (Valle et al. \cite{valle}) we found the
abundance curves to  systematically lie above the observations, we now
find very different  trends depending on the timing and efficiency of
the collisions.  This  difference reflects the changes in the SFR and
subsequent supernova  activity and the change in the cosmic ray
spallation production in the  altered diffuse gas.  The histories
depend sensitively on the timing of the  collisions, far more than for
the heavy elements, and argue against using  these species to
reconstruct a unique galactic history of star formation.  We show a
sample of single and multiple collision results in
Fig.~\ref{fig:all-light}.  In each simulation, we removed
$\approx$70\% of the diffuse gas present in the system at the onset
of the collision event, so these likely represent an upper
limit for the reduction of the final abundances. These simulations
suggest that after a few  strong collisions (and here we don't include
the distinction between a major and minor merging event since the
models are not dynamical) the final abundances can  be altered from
the \stdmdl\ by as much as a factor of three.

\subsection{Stripping Simulations}

In our treatment, stripping by an external medium and impulsive
collision  events are similar in their effects.  We have implemented
the effects of ram pressure from an intracluster medium by adopting a
characteristic timescale of 1.3 Gyr for the event and varying only the
initial time for the process using the Quilis et al.(\cite{quilis00})
mass loss profile  although their timescale is an order of magnitude
shorter.   Once again, because of the decline of the SFR, the timing
of the event is  important.  If the stripping begins early in the
history of the galaxy the  thin disk is stunted in its growth and the
metallicity is significantly  reduced compared to the \stdmdl.  The
histories adopted here are for single infall episodes (in the sense of the
galaxy falling in the cluster potential) and contrast with 
those  used for simulations of ram pressure stripping of Virgo cluster
galaxies by  Vollmer et al. (\cite{Vol01}) who find a
ram pressure history that is approximately Lorentzian in time with a
FWHM of about 200 Myr (see also Vollmer et al. \cite{Vol04}, and references
therein).  Again, our prescription is general, not specifically
dynamical, and can be adapted to any stripping history when combined
with a model for the galaxy motion within a realistic cluster
potential and background.

\begin{figure}
\centering
\includegraphics[width=8cm]{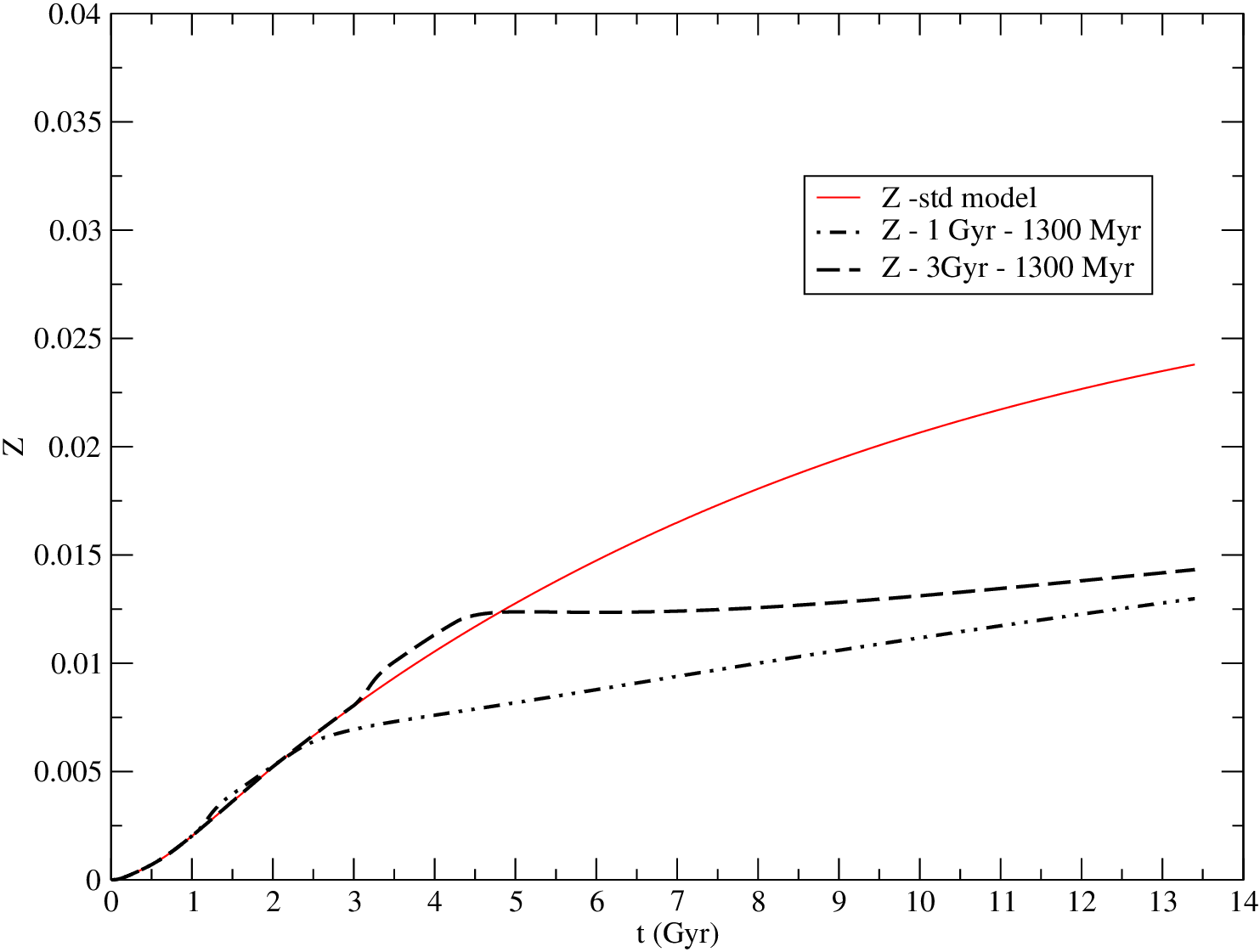}\hspace{10mm}
\includegraphics[width=8cm]{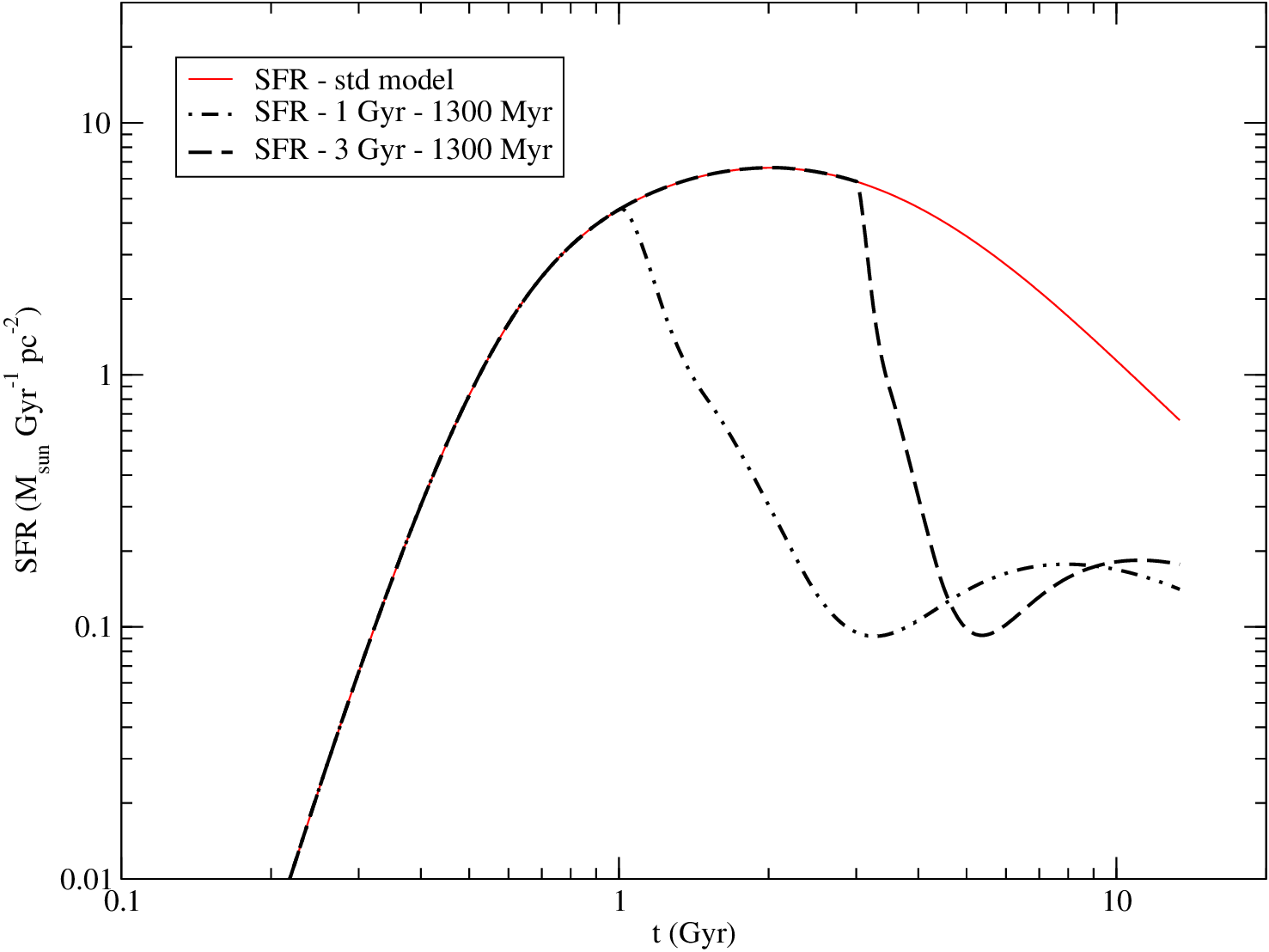}
\caption{Stripping simulations comparison with $t_{start}=1$ 
and $t_{start}=3$ Gyr, $\Delta
  t_{coll}=1.3$ Gyr using the Quilis et al. (\cite{quilis00})
  mass loss profile.  The residual gas fraction was 3\% after 
$\Delta t_{coll}$.}\label{fig:stripping-confronto-1-e-3gyr}
\end{figure}

We show our results for two simulations, one 
starting after the peak of the halo 
formation but before the thin disk starts to form and the other well after the 
start of star formation in the thin disk.  The final SFR is almost a 
factor of 10 lower than the \stdmdl\ and the metallicity is reduced by 
about 40\%.  The overall appearance of the system is similar to the 
collision case, the longer timescale for the continued gas removal having 
little additional effect compared to the impulsive loss since the SFR is 
so drastically reduced and the thin disk 
is severely suppressed (Fig.~\ref{fig:stripping-confronto-1-e-3gyr}).
The models yield behavior that resembles the results of de Mello,
Wikland, \& Maia (\cite{deM02}) who find that ``... repeated encounters,
experienced by galaxies in dense environments, removes gas from the
galaxies as well as leads to an inhibition of the formation of the
molecular gas from the atomic phase.  The similarities  in
star-formation efficiency of the dense environments and field galaxies
suggest that the physical processes controlling the formation of stars
from the molecular gas are local rather than global.''

For the observable HR diagram, we expect a range of morphologies.  For systems 
undergoing strong stripping, the giant branch should be diffuse with the bulk 
of the older stars being formed with lower metallicity 
and the higher metallicity stars showing an almost continuous distribution in 
color at any luminosity.  We will report the detailed simulations in a 
future paper but point out that in any galactic 
model the output from any of the 
simulations reported here can be inserted to predict color-magnitude diagrams 
as a function of time ({\it e.g.} Cignoni, Prada  Moroni \& 
Degl'Innocenti \cite{cig03}).

\section{Fully Open Systems with Combined Stripping and Re-accretion}

\begin{figure}
\vspace{5mm}
\centering
\includegraphics[width=8cm]{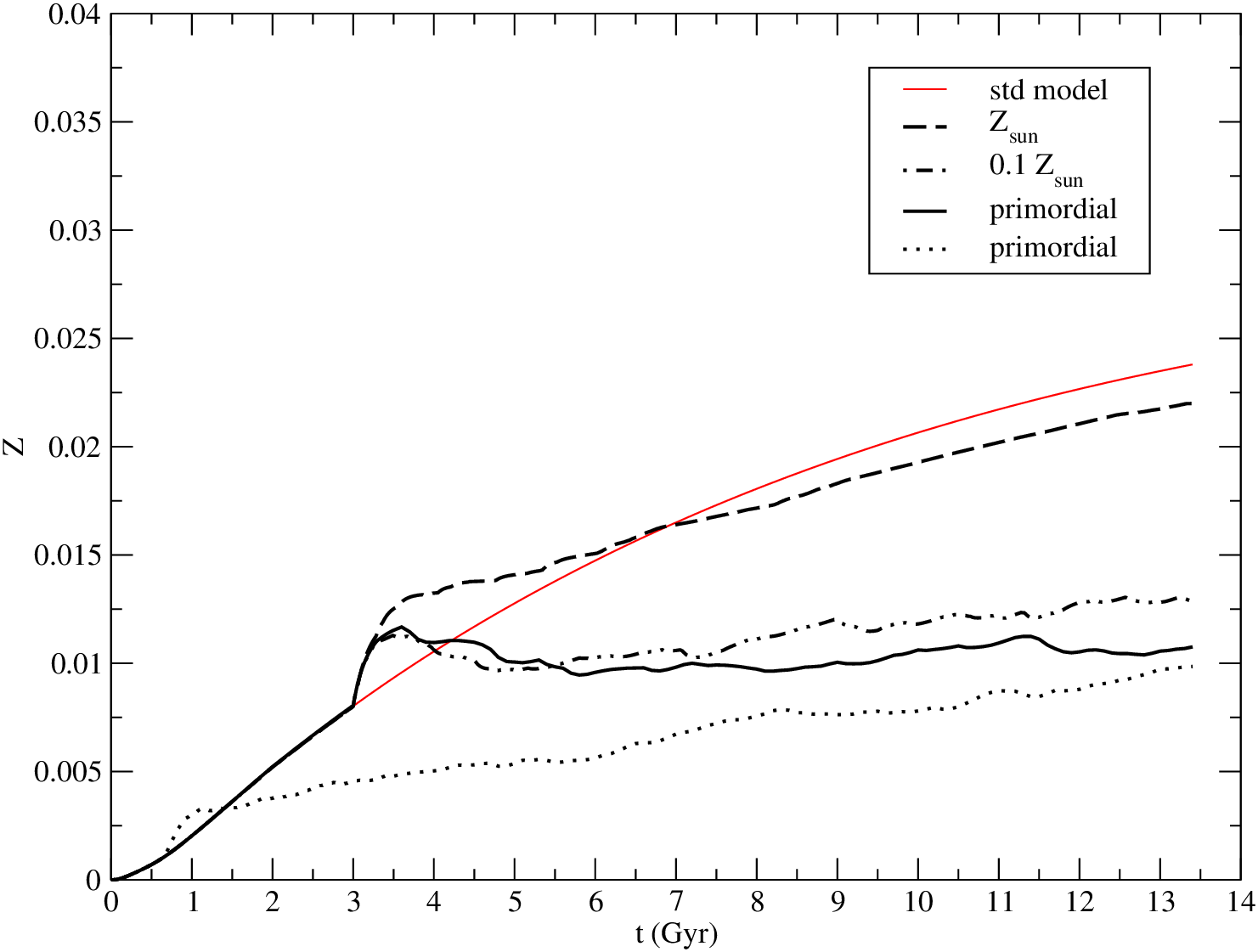}\hspace{10mm}
\includegraphics[width=8cm]{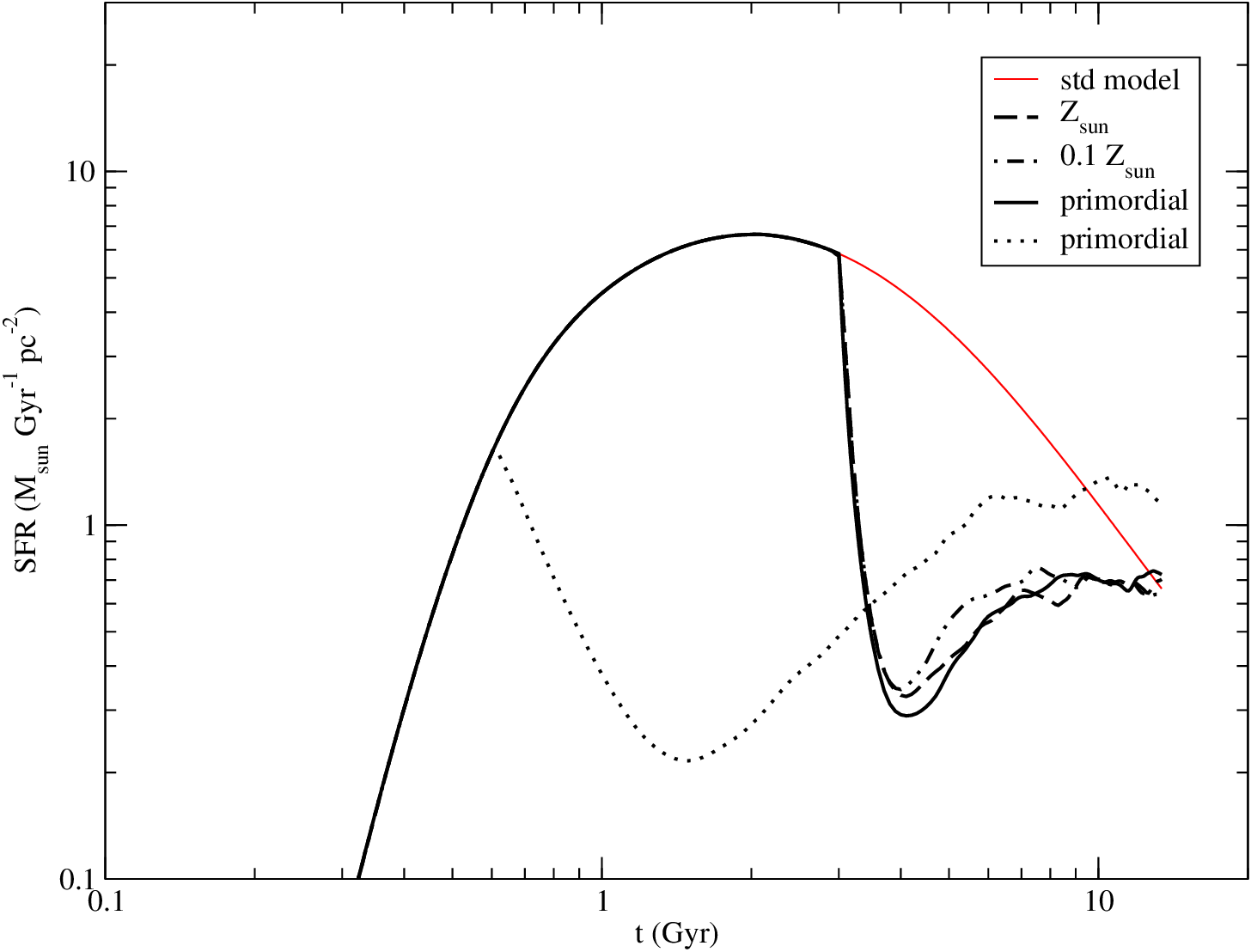}
\caption{Combined effects of collision and environmental infall.
The collision event was assumed to have $t_{start}=3$Gyr  
and $\Delta t_{coll}=50$ Myr (see text for details).
After the collision, the models assume stochastic infall, in the same 
way as in the Figs.~\ref{fig:open-accr-zsun},  
but the re-accretion after the end of the collision.  About 40\% of the
 total initial  mass was removed during the collision and then about
 10\% was restored during the subsequent re-accretion.  
Three principal cases of environmental infall are considered:
low metallicity gas ($0.1 \; Z_\odot$) (dot-dash), metal 
rich ambient gas ($Z_\odot$) (dash), and primordial material 
(black continuous).  One additional case is shown for re-accretion of 
primordial material (dotted line), starting the collision at an 
earlier epoch, 0.6 Gyr; for this last case, 25\% of the total 
initial mass is subsequently re-accreted.}
\label{fig:confronto-coll-accr-01ezsun}
\end{figure}

As a final simulation, we assume that both collisional stripping 
and accretion of ambient gas can occur (and, as Vollmer et
al. \cite{Vol01} point out, also re-accretion of stripped gas).  
The results are shown in Fig.~\ref{fig:confronto-coll-accr-01ezsun} 
for relatively low metallicity accretion (0.1 Z$_\odot$),
metal rich ambient gas ($Z_\odot$) and primordial material. 
Both the cases 0.1 Z$_\odot$ and Z$_\odot$ enriched relative 
to primordial abundances that provide an 
appropriate range consistent with observations of clusters.
The dominant effect comes from the removal of gas 
unless the infall rate is extremely high.  As we have shown in the previous 
sections, accretion doesn't simply dilute the abundances 
(Casuso \& Beckman \cite{casuso04}). That may be true in a 
closed box-type approach or one with a formally assumed SFR. We 
see instead that -- depending on the metallicity of the accreted
material and the timing and rates 
of the stripping and filling events -- 
the new gas powers further star formation and the 
resulting abundances are never 
reduced by more than about 40\%. The most divergent 
case is for high metallicity accretion 
where the effects of the collision are, 
as expected, completely erased although the reduced star 
formation over most of 
the system's history means the high metal stars are 
comparatively rare relative to 
the standard model.

We qualitatively anticipate one more interesting outcome 
of these combined ``stripping and filling'' scenarios.  Such post-collision 
systems may display anomalous 
color-magnitude diagrams with double red giant branches and strange 
age-metallicity relations.  For instance, the model in Fig.~\ref{fig:confronto-coll-accr-01ezsun} should 
show a metal rich population that is older than a rarer 
metal poor disk population.  This is just reversed from 
the \stdmdl\ and collisional stripping-only models.  
This can be understood by weighting $Z(t)$ by the time dependent SFR; 
although not a large effect in the models we computed, the 
fact that the $Z$ curve is not monotonic produces the gap in age and the 
lower metallicity for the more recently formed stars.  The greater the rate of 
resupply, the greater the contrast 
between the two populations.  Perhaps such 
color-magnitude exist for dwarf systems?  It would be interesting to test the 
picture for cases where the stripping is more devastating and the infall 
just now restarting.  This ``burst in reverse'' behavior would likely be 
enhanced in systems that experience stimulated bursts of star formation 
during the stripping event.  

\section{Summary}

Our aim here has been to display the behavior of a galactic model to a
wide range of physical processes that are expected to be important
during  some stage of Galactic history.  Because these calculations are
restricted to modeling a 
local region of the disk with multiple
vertical  stratification -- but without dynamics -- extended to a schematicized 
model for the whole galaxy, a detailed comparison
with the evolution  of the Solar neighborhood and the Galactic disk is
beyond our capabilities.  But  we have found a consistent behavior in
the presence of accretion and stripping  that resemble that seen in
both our and other galaxies.  From these simulations we conclude that
it is unlikely that one can uniquely reconstruct  the star formation
history of the Galaxy solely from the metallicity  distribution,
especially for the light elements.  Finally, we note  that it is now
computationally possible to merge our approach with  N-body SPH
modeling to follow the gas and population evolution self-consistently
during collisions and also for clusters. An important feature of the multiphase models, the separation of the gas into distinct diffusive and cloud phases, can also be extended to include enhanced stimulated star formation during collisions. This hybrid  modeling would
be able to follow the SFR and metallicity evolution  explicitly in
each individual galaxy within cosmological simulations. 

\begin{acknowledgements}
We thank Michele Cignoni, Federico Ferrini, Pepe Franco, 
Joachim K\"oppen, Francesco Palla, Jan Palou\v{s}, Pier Prada Moroni, 
and Nikos Prantzos for discussions. We also thank the referee, 
Jesper Sommer-Larsen, for valuable advice and criticisms.  
Giada Valle wish to thank 
Matteo Dell'Omodarme for remarkably patient and invaluable help and 
comments. This work was supported, in part, by COFIN2000 from MIUR.
\end{acknowledgements}

\end{document}